\documentclass[lettersize,journal]{IEEEtran}
\IEEEoverridecommandlockouts
\usepackage{geometry}
\usepackage{graphicx}
\usepackage{epstopdf}
\usepackage{cite}
\usepackage{amsmath,amssymb,amsfonts}
\usepackage{algorithmic}
\usepackage{float}
\usepackage{graphicx}
\usepackage{textcomp}
\usepackage{xcolor}
\usepackage{array}
\usepackage{multicol}
\usepackage{multirow}
\usepackage{makecell}
\allowdisplaybreaks
\usepackage{subcaption}
\usepackage[ruled,vlined]{algorithm2e}
\usepackage{bm}
\usepackage{booktabs}
\usepackage{array}
\usepackage{cancel}
\usepackage[normalem]{ulem}

\newtheorem{proposition}{Proposition}

\def\BibTeX{{\rm B\kern-.05em{\sc i\kern-.025em b}\kern-.08em
    T\kern-.1667em\lower.7ex\hbox{E}\kern-.125emX}}
\begin{document}

\title{Distributed Image Semantic Communication via Nonlinear Transform Coding
\author{
Yufei~Bo, Meixia~Tao, \textit{Fellow, IEEE}, and Kai~Niu, \textit{Member, IEEE}}
\thanks{Y. Bo, and M. Tao are with the Department of Electronic Engineering and Shanghai Key Laboratory of Digital Media Processing and Transmission, Shanghai Jiao Tong University, Shanghai, 200240, China (emails: \{boyufei01, mxtao\}@sjtu.edu.cn).}
\thanks{Kai Niu is with the Key Laboratory of Universal Wireless Communications, Ministry of Education, Beijing University of Posts and Telecommunications, Beijing 100876, China, and also with the Peng Cheng Laboratory, Shenzhen 518066, China (e-mail: niukai@bupt.edu.cn).}
\thanks{The code of this work will be released at https://github.com/SJTUmxtao/D-NTSCC after publication.}
\thanks{A preliminary version of this work was accepted at the IEEE ICC Workshop, Montreal, Canada, Jun. 2025\cite{conference_ver}.}}
\maketitle

\begin{abstract}

This paper investigates distributed source-channel coding for correlated image semantic transmission over wireless channels. 
In this setup, correlated images at different transmitters are separately encoded and transmitted through dedicated channels for joint recovery at the receiver. 
We propose a general approach for distributed image semantic communication that applies to both separate source and channel coding (SSCC) and joint source-channel coding (JSCC).
Unlike existing learning-based approaches that implicitly learn source correlation in a purely data-driven manner, our method leverages nonlinear transform coding (NTC) to explicitly model source correlation from both probabilistic and geometric perspectives. 
A joint entropy model approximates the joint distribution of latent representations to guide adaptive rate allocation, while a transformation module aligns latent features for maximal correlation learning at the decoder.
We implement this framework as D-NTSC for SSCC and D-NTSCC for JSCC, both built on Swin Transformers for effective feature extraction and correlation exploitation.
Variational inference is employed to derive principled loss functions that jointly optimize encoding, decoding, and joint entropy modeling.
Extensive experiments on real-world multi-view datasets demonstrate that D-NTSC and D-NTSCC outperform existing distributed SSCC and distributed JSCC baselines, respectively, achieving state-of-the-art performance in both pixel-level and perceptual quality metrics.
\end{abstract}


\section{Introduction}

Semantic communication, or \textit{task-oriented communication}, has emerged as a promising paradigm for future 6G networks\cite{2024qin, gunduz2022beyond}.
It aims to transmit only the information that is essential for accomplishing specific tasks at the receiver.
By leveraging deep learning, semantic communication systems replace traditional hand-crafted communication modules by neural network (NN)-based counterparts.
Most existing semantic communication systems follow either a separate source and channel coding (SSCC) architecture or a joint source-channel coding (JSCC) architecture.
The ones with SSCC adopt a modular design, where source compression and channel coding are optimized independently \cite{2023huang}. 
In contrast, the ones with JSCC jointly optimize source and channel coding in an end-to-end manner, taking into account source distribution, channel distribution, as well as receiver tasks\cite{jscc, jscc2, ntscc, mywork}.
Notably, compared with conventional Shannon-type communication, semantic communication offers significant improvements in transmission efficiency and downstream task performance across various sources \cite{jscc, 2021speech, video} and tasks \cite{classification, detection}.
While early studies primarily focused on point-to-point communication \cite{2023huang, jscc, jscc2, ntscc, mywork}, recent research has extended semantic communication to multi-user scenarios\cite{yilmaz2023distributed, mywork2, multiuservqa}, including multiple access channels\cite{yilmaz2023distributed} and broadcast settings\cite{mywork2}.

This work investigates multi-user semantic communications from distributed transmitters with correlated image sources to a common receiver. 
The goal at the receiver is to reconstruct the images with the highest fidelity at minimum channel usage. 
A typical application scenario is multi-view surveillance systems, where multiple separately located cameras with probably overlapping fields of view capture images and transmit them to a center node for joint reconstruction. 
Traditionally, this problem is formulated as distributed source coding (DSC).
As demonstrated by the Slepian-Wolf (SW) theorem\cite{slepian1973noiseless}, separate encoding and joint decoding of correlated sources can achieve the same compression rate as joint encoding-decoding under lossless compression. 
This theorem has then been extended to the lossy case by Berger\cite{berger1978multiterminal} and Tung\cite{tung1978multiterminal}.
Building upon these information theoretical results, many practical DSC schemes have been proposed, such as \cite{pradhan2003distributed, yang2008multiterminal}.  
These approaches, designed for binary sources with simple correlation models, suggest the use of a linear parity-check channel code, such as low-density parity-check (LDPC) code.
By partitioning the binary sources into cosets indexed by \textit{syndromes} of a channel code, they are able to achieve the SW bound.

Recently, deep learning-based DSC methods have been developed to exploit correlations directly from image data\cite{diao2020drasic, ldmic}.
The authors in \cite{diao2020drasic} propose a recurrent autoencoder-based architecture with multiple distributed encoders and a single decoder to implicitly learn the correlations among images.
The work in \cite{ldmic} further proposes a joint context transfer module at the decoder, leveraging the cross-attention (CA) mechanism to capture inter-view correlations.
While these methods focus on optimizing source compression, recent advances have also focused on distributed JSCC\cite{wang2022distributed, li2024content}, where channel conditions are integrated into the end-to-end design.
The work \cite{wang2022distributed} employs an autoencoder-based architecture with two lightweight edge encoders and a central decoder, introducing a signal-to-noise-ratio (SNR)-aware CA module at the receiver for enhanced feature interaction.
Additionally, the work \cite{li2024content} introduces a multi-layer CA mechanism to fuse image features across different pixel levels for better correlation exploitation.
In \cite{dong2024robust}, the authors extend this problem to fading channels with imperfect CSI and propose a robust distributed JSCC method.
However, existing approaches rely on implicitly learning source correlation in a purely data-driven manner, without incorporating prior knowledge of correlation structures.
This motivates the development of more principled strategies to exploit such correlation.

In this paper, we propose a general approach for distributed semantic communication that leverages nonlinear transform coding (NTC) to exploit source correlation explicitly through joint distribution modeling.
We develop two schemes, D-NTSC and D-NTSCC, for SSCC and JSCC, respectively.
Specifically, in both cases, multiple correlated image sources are separately encoded by nonlinear analysis transforms into latent representations.
These latent representations are then either quantized and entropy-coded (in the SSCC case), or mapped directly to channel input symbols  through JSCC encoders (in the JSCC case).
The resulting signals are transmitted over independent noisy channels to a common receiver.
Their transmission rates are determined in an adaptive manner based on the entropy of the latent representations, which is approximated by a learned joint entropy model.
At the receiver, the latent representations are recovered and jointly decoded using a joint nonlinear synthesis transform, where each image utilizes the other noisy latent representations as side information.

The contributions of this paper can be summarized as follows.
\begin{itemize}
    \item We propose a general NTC-based framework for distributed image semantic communication, applicable to both SSCC and JSCC scenarios, and develop two corresponding schemes, D-NTSC and D-NTSCC.
    This framework explicitly models source correlation from both probabilistic and geometric perspectives. The former is achieved through joint entropy modeling based on NTC, while the latter is addressed via spatial alignment at the receiver.
    In addition, this framework is built upon the Swin Transformer architecture, which facilitates more effective global and local feature extraction, as well as correlation exploitation.
    
    \item From the probabilistic perspective, we develop a joint entropy model to approximate the joint distribution of latent representations, which more accurately captures their dependencies and improves rate-distortion performance.
    Furthermore, we derive principled loss functions for both D-NTSC and D-NTSCC based on variational inference, which jointly optimize the encoder, decoder, and entropy model to ensure that the learned joint entropy model closely matches the true joint distribution.
    
    \item From the geometric perspective, we propose a transformation module at the receiver to locate and transform the most relevant side information in the other latent representations to an expected pose, ensuring maximal correlation utilization in subsequent layers.
    
    \item Extensive experiments on real-world multi-view datasets validate the advantages of the proposed D-NTSC and D-NTSCC schemes.
    D-NTSC outperforms existing distributed source coding baselines across various bit rates, while D-NTSCC achieves superior performance over state-of-the-art distributed JSCC methods under varying transmission rates and channel conditions, attaining the best results in both pixel-wise and perceptual metrics.
    Ablation studies demonstrate that the joint entropy model improves rate-distortion performance, while the transformation module effectively aligns the latent representations.
    
\end{itemize}

The remainder of this paper is organized as follows.
Section II reviews the basic principles of NTC.
Section III presents the system model and overall architecture of the proposed D-NTSC and D-NTSCC schemes.
In Section IV, we formulate the optimization objectives based on variational inference.
Section V details the architecture of key NN modules.
Comprehensive experiments are conducted in Section VI to evaluate the performance of the proposed methods.
Finally, Section VII concludes the paper.

Throughout this paper, bold lowercase letters (\emph{e.g.}, $\mathbf{x}$) and bold uppercase letters (\emph{e.g.}, $\mathbf{X}$) denote vectors and matrices, respectively. The identity matrix of size $k \times k$ is denoted by $\mathbf{I}_{k \times k}$. The entropy of a random variable $X$ is denoted by $H(X)$, and its statistical expectation is written as $\mathbb{E}[X]$.
The Kullback–Leibler (KL) divergence between two probability distributions $q$ and $p$ is denoted by $D_{KL}[q||p]$.
We denote the circularly symmetric complex Gaussian distribution with mean $\mu$ and variance $\sigma^2$ as $\mathcal{CN}(\mu, \sigma^2)$. 
For probability notation, $p_x$ refers to the probability density function (PDF) of a continuous random variable $x$, and $P_{\bar{x}}$ denotes the probability mass function (PMF) of a discrete random variable $\bar{x}$.

\section{Backgrond: Nonlinear Transform Coding}

Classical image coding methods generally adopt the following procedure: first, the image is transformed into a latent representation, followed by quantization in this transformed space. The resulting discrete vector is further compressed using entropy coding methods, such as arithmetic coding, to create a bitstream for further processing.
NTC is an image coding technique that applies nonlinear transforms to an image before quantization and entropy coding\cite{ntc}, in contrast to linear transforms used in JEPG\cite{jpeg}, BPG\cite{bellard2014bpg}, \textit{etc.}.

\begin{figure}[t]
    \centering
    \includegraphics[width=0.45\textwidth]{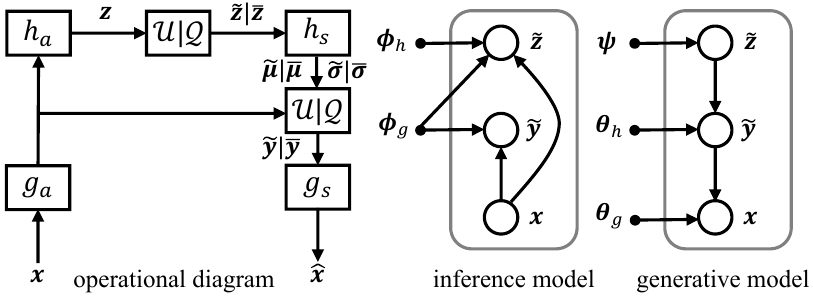}
    \caption{\small{
    Left: Operational diagram of NTC. 
    The label $\mathcal{U|Q}$ denotes the use of uniform noise during training and actual quantization during testing.
    Right: Graphical representation of the analysis transform as an inference model, and the synthesis transform as a generative model in NTC. 
    Hollow nodes indicate random variables, solid nodes represent parameters, and arrows denote conditional dependencies.}}
    \label{graphical ntc}
    \vspace{-0.5cm}
\end{figure}

Specifically, as shown on the left side of Fig.~\ref{graphical ntc}, the encoder in NTC maps an image $\mathbf{x}$ to a latent representation $\mathbf{y}$ using a nonlinear analysis transform $g_a(\mathbf{x};\boldsymbol{\phi}_g)$, which is quantized to $\mathbf{\bar{y}}$.
The quantized sequence $\mathbf{\bar{y}}$ is then entropy coded for transmission, under the assumption of an error-free channel to ensure perfect recovery at the decoder.
The decoder reconstructs the image $\mathbf{\hat{x}}$ from $\mathbf{\bar{y}}$ via a nonlinear synthesis transform $g_s(\mathbf{\bar{y}};\boldsymbol{\theta}_g)$.
The nonlinear transforms $g_a$ and $g_s$ are typically parameterized by NNs due to their universal function approximation capabilities, with $\boldsymbol{\phi}_g$ and $\boldsymbol{\theta}_g$ denoting their respective parameters.
To enable entropy coding, NTC further establishes an entropy model to approximate the distribution of the latent representation $\mathbf{y}$.
To this end, an auxiliary latent variable $\mathbf{z}$, referred to as a \textit{hyperprior}, is introduced to represent the dependencies within $\mathbf{y}$.
An analysis transform $h_a(\mathbf{y};\boldsymbol{\phi}_h)$ maps $\mathbf{y}$ to $\mathbf{z}$.
Conditioned on $\mathbf{z}$, the elements of $\mathbf{y}$ are assumed to be independent and modeled as Gaussian variables.
The corresponding means and standard deviations are predicted by applying a synthesis transform $h_s(\mathbf{\bar z};\boldsymbol{\theta}_h)$ to $\mathbf{\bar z}$, where $\mathbf{\bar z}$ denotes the quantized version of $\mathbf{z}$.
During training, quantization is relaxed by adding uniform noise to avoid the zero-gradient problem.
The relaxed versions of $\mathbf{\bar y}$ and $\mathbf{\bar z}$ are denoted by $\mathbf{\tilde y}$ and $\mathbf{\tilde z}$, respectively.

\begin{figure*}
    \centering
    \includegraphics[scale=0.52]{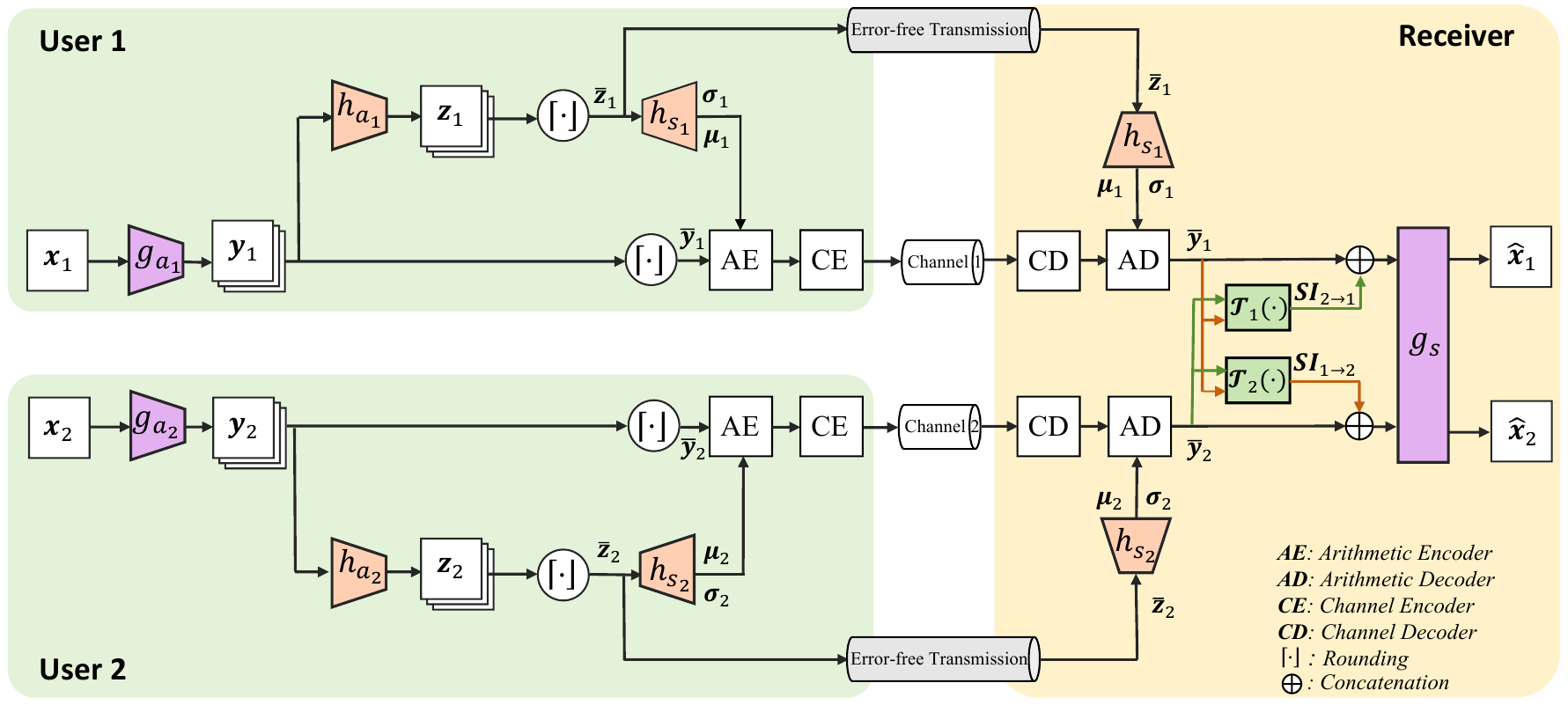}
    \caption{Overall architecture of the proposed D-NTSC scheme.}
    \label{system_model_ntsc}
    \vspace{-0.6cm}
\end{figure*}

The optimization problem for NTC is formulated as a rate-distortion trade-off, where the code rate $R$ is defined as the expected length of the quantized sequence, determined by the nonlinear transform and the entropy model.
This framework can also be interpreted as a variational autoencoder (VAE), where the analysis transform at the encoder corresponds to the inference model, and the synthesis transform at the decoder corresponds to the generative model\cite{hyperprior}, as shown on the right side of Fig.~\ref{graphical ntc}.
The objective is to approximate the true posterior of the inference model $p(\mathbf{\tilde y}, \mathbf{\tilde z} |\mathbf{x})$ using an NN-parameterized variational density $q(\mathbf{\tilde y}, \mathbf{\tilde z}|\mathbf{x})$ by minimizing their expected KL divergence
\begin{equation}
    \mathbb{E}_{p_{\mathbf{x}}}D_{KL}[q(\mathbf{\tilde y}, \mathbf{\tilde z}|\mathbf{x})||p(\mathbf{\tilde y}, \mathbf{\tilde z} |\mathbf{x})].
\end{equation}
By more closely adapting to the source distribution, NTC has become competitive with the best linear transform codecs for images, and has outperformed them in rate-distortion performance under perceptual quality metrics such as MS-SSIM.

While the above works\cite{ntc, hyperprior} focus on source coding, work \cite{ntscc} combines NTC with deep JSCC, proposing a JSCC architecture called nonlinear transform source-channel coding (NTSCC).
Unlike other deep JSCC methods, NTSCC first transforms the image into a latent representation, then applies deep JSCC in this latent domain.
The introduction of the entropy model enables the deep JSCC codec to adopt an appropriate coding scheme based on the distribution of the latent representation.

NTC, along with its JSCC extension, shows that the encoding that matches the source distribution leads to significant coding gains.
We extend this principle to the distributed image semantic communication scenario by proposing a framework that characterizes the \textit{joint distribution} of correlated sources.
The encoding and transmission schemes for multiple users are adapted to the joint source distribution, thus improving the performance across all the users.



\section{System Model and Overall Architecture}

This section outlines the system models of our proposed distributed coding framework, including the SSCC-based D-NTSC and the JSCC-based D-NTSCC schemes.

Overall, we consider semantic communications in a distributed setting, where two physically separated terminals send correlated but not necessarily independently distributed image sources to a common receiver through independent channels for joint reconstruction.
Here, two source terminals are considered for simplicity. 
The extension to more than two sources is straightforward. 
Let the two images be denoted as $\mathbf{x}_1, \mathbf{x}_2\in \mathbb{R}^{C\times H\times W}$, where $C$, $H$, and $W$ represent the channel, height, and width of the images, respectively.
In practice, $\mathbf{x}_1$ and $\mathbf{x}_2$ can be a pair of stereo images taken from two cameras with probably overlapping fields of view in a multi-camera surveillance system.

\subsection{System Model of D-NTSC}

The overall architecture of the proposed D-NTSC is shown in Fig.~\ref{system_model_ntsc}. 
Specifically, each transmitter $i$, for $i\in \{1,2\}$, first employs an NN-based analysis transform $g_{a_i}$, parameterized by $\boldsymbol{\phi}_{g_i}$, to map $\mathbf{x}_i$ into a latent representation $\mathbf{y}_i\in\mathbb{R}^{N_y}$.
Here $N_y$ denotes the length of $\mathbf{y}_i$.
Uniform scalar quantization $\lfloor \cdot \rceil$ (\emph{i.e.}, rounding to integers) is then applied to the vector $\mathbf{y}_i$ to obtain $\mathbf{\bar{y}}_i$, which is further compressed using arithmetic coding to create a bitstream.
Finally, channel coding is employed to ensure an error-free recovery of $\mathbf{\bar{y}}_i$ at the receiver.

The receiver first recovers $\mathbf{\bar{y}}_1$ and $\mathbf{\bar{y}}_2$ through channel decoding and arithmetic decoding.
Then, an NN-based synthesis transform $g_s$ with parameters $\boldsymbol{\theta}_{g}$ performs joint decoding, reconstructing $\mathbf{\hat{x}}_1$ and $\mathbf{\hat{x}}_2$ from $\mathbf{\bar{y}}_1$ and $\mathbf{\bar{y}}_2$.
During this process, the two quantized latent representations serve as side information for the image reconstruction of each other.
Given the perspective differences in the latent representations, we introduce a transformation module to warp the side latent representation toward an expected pose, thereby facilitating more effective correlation utilization in subsequent NN layers.
Specifically, $\mathcal{T}_1(\cdot)$ spatially aligns $\mathbf{\bar y}_2$ with $\mathbf{\bar y}_1$. 
The transformed $\mathbf{\bar y}_2$, namely, $\mathcal{T}_1(\mathbf{\bar y}_2)$, is denoted as $\mathbf{SI}_{2\rightarrow 1}$.
The same applies to $\mathcal{T}_2(\cdot)$.
Details of this module shall be introduced in Section V-C.
The main latent representation is then concatenated with the transformed side representation and fed into the synthesis transform $g_s$.
Therefore, the decoding process can be expressed as
\begin{align}
    \mathbf{\hat x}_1 &= g_s(\mathbf{\bar y}_1|\mathcal{T}_1(\mathbf{\bar y}_2)),\\
    \mathbf{\hat x}_2 &= g_s(\mathbf{\bar y}_2|\mathcal{T}_2(\mathbf{\bar y}_1)).
\end{align}


To enable arithmetic coding, a hyperprior variable $\mathbf{z}_i\in\mathbb{R}^{N_z}$ is introduced to characterize the PDF of $\mathbf{y}_i$, where $N_z\ll N_y$.
A parametric analysis transform $h_{a_i}$, with parameters $\boldsymbol{\phi}_{h_i}$, maps $\mathbf{y}_i$ to $\mathbf{z}_i$.
The hyperprior represents the dependencies among the elements of $\mathbf{y}_i$, allowing them to be assumed independent given $\mathbf{z}_i$\cite{hyperprior}.
Additionally, it summarizes the means and standard deviations of $\mathbf{y}_i$, and is quantized, compressed, and transmitted to the receiver.
Specifically, each element $y_i^j$ of $\mathbf{y}_i$, for $j=1,...,N_y$, is modeled as a Gaussian variable with mean $\mu_i^j$ and standard deviation $\sigma_i^j$, both estimated by applying a parametric synthesis transform $h_{s_i}$ to the quantized hyperprior $\mathbf{\bar{z}}_i$.
Note that, to be able to use gradient descent methods for the optimization of the network, quantization is approximated by adding uniform noise during training and replaced with actual rounding during inference.
We denote the approximated version of $\mathbf{\bar{y}}_i$ and $\mathbf{\bar{z}}_i$ as $\mathbf{\tilde{y}}_i$ and $\mathbf{\tilde{z}}_i$, respectively.
Therefore, the probability of $\mathbf{\tilde{y}}_i$ given $\mathbf{\tilde{z}}_i$ is obtained by convolving the probability of $\mathbf{y}_i$ with a standard uniform density, which can be written as
\begin{equation}
p(\mathbf{\tilde{y}}_i|\mathbf{\tilde{z}}_i)=\prod_j \left ( \mathcal{N}(\mu_i^j, \sigma_i^j)*\mathcal{U}(-\frac{1}{2}, \frac{1}{2}) \right )(\tilde{y}_i^j),
\label{py}
\end{equation}
with
\begin{equation}
    \boldsymbol{\sigma}_i, \boldsymbol{\mu}_i =h_{s_i}(\mathbf{\tilde{z}}_i;\boldsymbol{\theta}_{h_{i}}),
\end{equation}
where $*$ denotes the convolution operation, and $\boldsymbol{\theta}_{h_{i}}$ encapsulates the parameters of $h_{i}$.
Therefore, the expected code length required to transmit $\mathbf{y}_i$ is given by 
\begin{equation}
    R_{y_i}=\mathbb{E}_{p_{\mathbf{x}_i}}[-\log_2 p(\mathbf{\bar{y}}_i|\mathbf{\bar{z}}_i)] \ \ \ bits.
\end{equation}


Moreover, since the hyperpriors contain information about the code rate, they must also be available at the receiver for decoding.
Assuming error-free transmission, we estimate their transmission cost using a joint density model $p(\mathbf{\tilde{z}}_1, \mathbf{\tilde{z}}_2|\boldsymbol{\psi})$, where $\boldsymbol{\psi}$ denotes the distribution parameters.
Details of this joint density model are provided in Section IV-C.
Therefore, their joint entropy can be expressed as
\begin{equation}
    H(\mathbf{\bar{z}}_1, \mathbf{\bar{z}}_2) = \mathbb{E}_{p(\mathbf{\bar{z}}_1, \mathbf{\bar{z}}_2|\boldsymbol{\psi})}[-\log_2 p(\mathbf{\bar{z}}_1, \mathbf{\bar{z}}_2|\boldsymbol{\psi})] \ \ \ bits,
\end{equation}
and the total code rate $R_i$ for User $i$ is
\begin{equation}
    R_i = R_{y_i} + \frac{1}{2}H(\mathbf{\bar{z}}_1, \mathbf{\bar{z}}_2) \ \ \ bits.
    \label{code rate}
\end{equation}


\subsection{System Model of D-NTSCC}

\begin{figure*}
    \centering
    \includegraphics[scale=0.52]{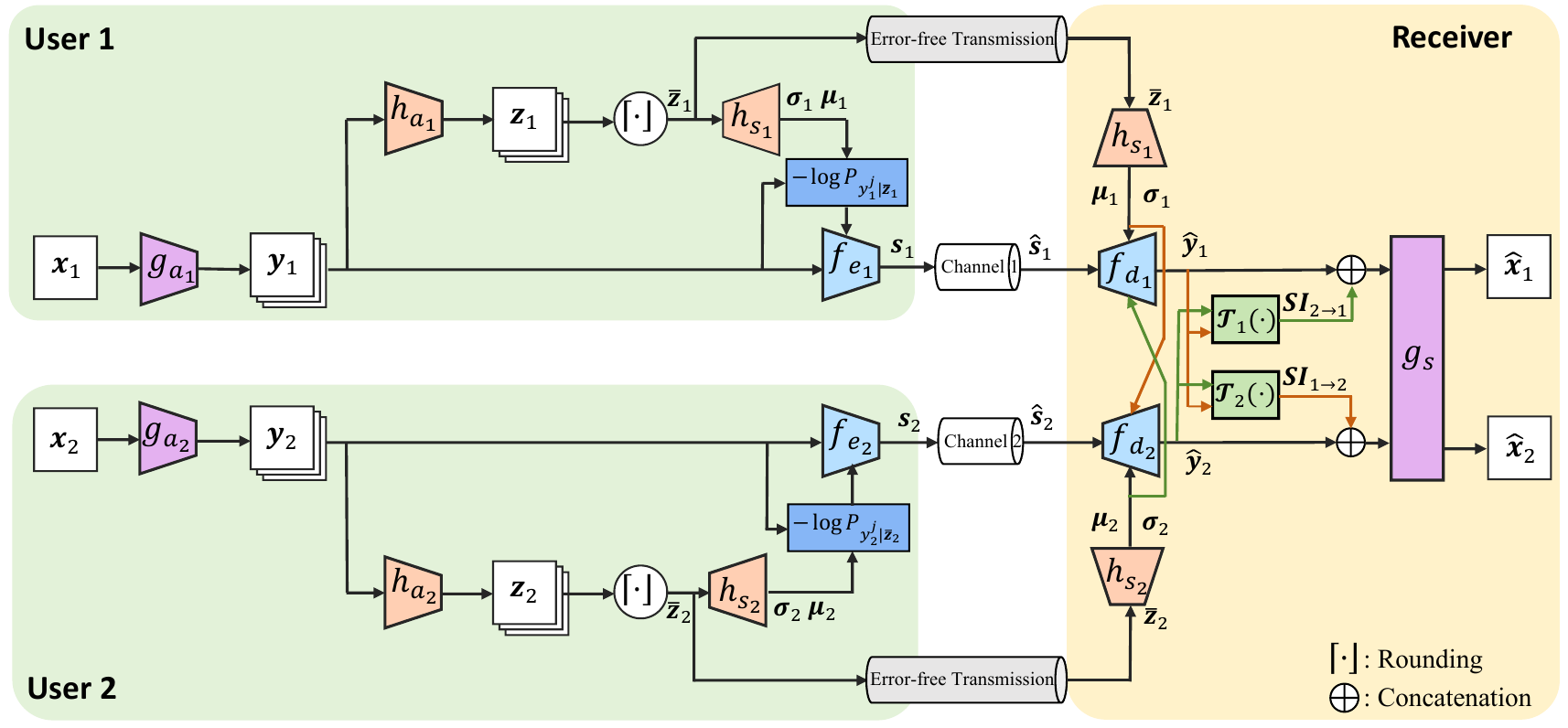}
    \caption{Overall architecture of the proposed D-NTSCC scheme.}
    \label{system_model_ntscc}
    \vspace{-0.6cm}
\end{figure*}

Under the same setup, the overall architecture of the proposed D-NTSCC is shown in Fig.~\ref{system_model_ntscc}.
Unlike the D-NTSC scheme where the source coding and channel coding are optimized separately, D-NTSCC focuses on an integrated design of source and channel coding.
The quantization, arithmetic coding and channel coding in D-NTSC are replaced by a JSCC encoder, while channel decoding and arithmetic decoding are replaced by a JSCC decoder.
Specifically, an analysis transform $g_{a_i}$ first extracts the latent representation $\mathbf{y}_i$, for $i\in \{1,2\}$.
Then, instead of quantization, a JSCC encoder $f_{e_i}$, parameterized by $\boldsymbol{\phi}_{f_i}$, encodes $\mathbf{y}_i$ into the channel input $\mathbf{s}_i$, which is reshaped into a complex-valued vector $\mathbf{s}_i\in\mathbb{C}^{n_i}$ for transmission.
Here, $n_i$ denotes the number of channel use of User $i$. 
Each element of $\mathbf{s}_i$ is subject to an average transmit power constraint 
$P$, which is assumed to be the same for both transmitters.

Furthermore, to achieve adaptive rate transmission, the transmission rate of $\mathbf{s}_i$ is constrained proportionally to the rate of $\mathbf{y}_i$, which is similarly characterized using hyperprior variables, as in D-NTSC.
The code rate of $\mathbf{y}_i$ is given by $-\log_2 p(\mathbf{\tilde{y}}_i|\mathbf{\tilde{z}}_i)$, with $p(\mathbf{\tilde{y}}_i|\mathbf{\tilde{z}}_i)$ defined in \eqref{py}.
Note that although $\mathbf{y}_i$ is not quantized in D-NTSCC, to ensure training stability, its probability distribution is also modeled by convolving it with a standard uniform density.
The hyperpriors are transmitted in the same manner as in the D-NTSC scheme.
And the overall transmission rate for User $i$ in \textit{channel uses / image dimension} is defined as
\begin{align}
    r_i = \frac{1}{CHW}\left [n_i + \frac{1}{2\times \mathcal{C}} H(\mathbf{\bar{z}}_1, \mathbf{\bar{z}}_2) \right ], 
\label{def r}
\end{align}
where $\mathcal{C}$ denotes the channel capacity, and the term $\frac{1}{2\times \mathcal{C}} H(\mathbf{\bar{z}}_1, \mathbf{\bar{z}}_2)$ corresponds to the channel use of transmitting the hyperpriors using capacity-achieving channel codes.

We consider a dedicated and independent additive white Gaussian noise (AWGN) channel for each transmitter.
The received vector is expressed as
$ \mathbf{\hat s}_i = \mathbf{s}_i + \mathbf{n}_i $, 
where $\mathbf{n}_i\sim \mathcal{CN}(\mathbf{0}, \epsilon^2_i\mathbf{I}_{n\times n})$ is the i.i.d. complex Gaussian noise with zero mean and variance $\epsilon^2_i$.
And we assume equal noise variance for both users, \emph{i.e.,} $\epsilon^2_1=\epsilon^2_2=\epsilon^2$.
Thus, the two channels have the same SNR, defined as $\frac{P}{\epsilon^2}$.

At the receiver, the received vector $\mathbf{\hat s}_i$ first passes through a JSCC decoder $f_{d_i}$, parameterized by $\boldsymbol{\theta}_{f_i}$, to recover $\mathbf{\hat{y}}_i$ with the assistance of $\mathbf{\bar{z}}_1$ and $\mathbf{\bar{z}}_2$.
The subsequent joint decoding follows the same procedure as in the D-NTSC scheme.
Specifically, the recovered latent representations $\mathbf{\hat{y}}_1$ and $\mathbf{\hat{y}}_2$ are concatenated with the transformed side information $\mathbf{SI}_{2\rightarrow 1}$ and $\mathbf{SI}_{1\rightarrow 2}$, respectively, and sent into the synthesis transform $g_s$ to obtain the reconstructed images $\mathbf{\hat{x}}_1$ and $\mathbf{\hat{x}}_2$.

\section{Optimization with Variational Inference}

In this section, we first derive the loss functions for D-NTSC and D-NTSCC using variational inference, as detailed in Sections IV-A and IV-B, respectively. 
We then develop a joint probabilistic model for the hyperprior variables in Section IV-C.

\begin{figure}[t]
    \centering
    \includegraphics[width=0.45\textwidth]{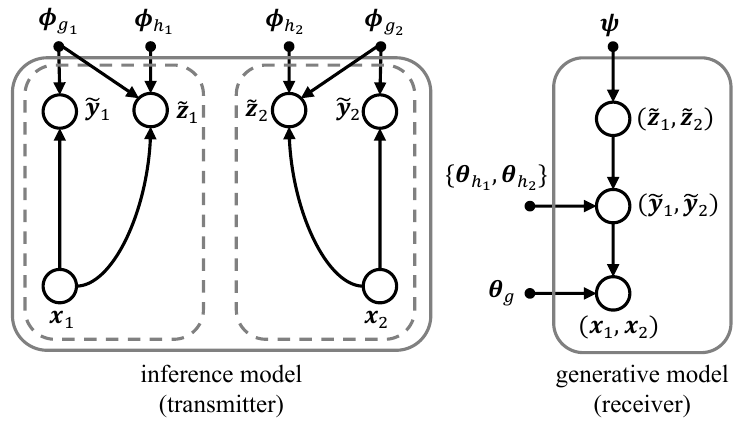}
    \caption{\small{Graphical representation of distributed analysis transforms as an inference model, and the joint synthesis transform as a generative model for D-NTSC.}}
    \label{graphical ntsc}
    \vspace{-0.6cm}
\end{figure}

\subsection{Variational Inference for D-NTSC}

\begin{figure*}
    \begin{align}
        &\ \ \ \ \ \mathbb{E}_{p_{\mathbf{x}_1,\mathbf{x}_2}}D_{KL}[q(\mathbf{\tilde{y}}_1,\mathbf{\tilde{y}}_2,\mathbf{\tilde{z}}_1,\mathbf{\tilde{z}}_2|\mathbf{x}_1,\mathbf{x}_2)||p(\mathbf{\tilde{y}}_1,\mathbf{\tilde{y}}_2,\mathbf{\tilde{z}}_1,\mathbf{\tilde{z}}_2|\mathbf{x}_1,\mathbf{x}_2)]\nonumber\\
        &=\mathbb{E}_{p_{\mathbf{x}_1,\mathbf{x}_2}}\mathbb{E}_{q_{\mathbf{\tilde{y}}_1,\mathbf{\tilde{y}}_2,\mathbf{\tilde{z}}_1,\mathbf{\tilde{z}}_2|\mathbf{x}_1,\mathbf{x}_2}} \Bigl\{ 
        \underbrace{-\log p(\mathbf{x}_1,\mathbf{x}_2|\mathbf{\tilde{y}}_1,\mathbf{\tilde{y}}_2,\mathbf{\tilde{z}}_1,\mathbf{\tilde{z}}_2)}_{\text{weighted distortion}} 
        -\underbrace{\log p(\mathbf{\tilde{y}}_1|\mathbf{\tilde{z}}_1)-\log p(\mathbf{\tilde{y}}_2|\mathbf{\tilde{z}}_2)-\log p(\mathbf{\tilde{z}}_1,\mathbf{\tilde{z}}_2)}_{\text{rate}} 
       \Bigr\} + \text{const}
        \label{sscc kl}
    \end{align}
    \vspace{-0.6cm}
\end{figure*}

\begin{figure*}
    \begin{align}
        &\ \ \ \ \ \mathbb{E}_{p_{\mathbf{x}_1,\mathbf{x}_2}}D_{KL}[q(\hat{\mathbf{s}}_1,\hat{\mathbf{s}}_2,\tilde{\mathbf{z}}_1,\tilde{\mathbf{z}}_2|\mathbf{x}_1,\mathbf{x}_2)||p(\hat{\mathbf{s}}_1,\hat{\mathbf{s}}_2,\tilde{\mathbf{z}}_1,\tilde{\mathbf{z}}_2|\mathbf{x}_1,\mathbf{x}_2)]\nonumber\\
        &=\mathbb{E}_{p_{\mathbf{x}_1,\mathbf{x}_2}}\mathbb{E}_{q_{\hat{\mathbf{s}}_1,\hat{\mathbf{s}}_2,\tilde{\mathbf{z}}_1,\tilde{\mathbf{z}}_2|\mathbf{x}_1,\mathbf{x}_2}} \Bigl\{ 
        \underbrace{-\log p(\mathbf{x}_1,\mathbf{x}_2|\hat{\mathbf{s}}_1,\hat{\mathbf{s}}_2,\tilde{\mathbf{z}}_1,\tilde{\mathbf{z}}_2)}_{\text{weighted distortion}} 
        -\underbrace{\log p(\hat{\mathbf{s}}_1|\tilde{\mathbf{z}}_1)-\log p(\hat{\mathbf{s}}_2|\tilde{\mathbf{z}}_2)-\log p(\tilde{\mathbf{z}}_1,\tilde{\mathbf{z}}_2)}_{\text{rate}} 
       \Bigr\} + \text{const}''
        \label{kl}
    \end{align}
    \hrulefill
    \vspace{-0.6cm}
\end{figure*}

As is stated beforehand, the optimization problem of NTC is typically formulated as a VAE\cite{hyperprior}.
Fig.~\ref{graphical ntsc} illustrates the corresponding VAE model for D-NTSC, where the analysis transforms at the transmitters are the inference model, and the joint synthesis transform at the receiver corresponds to the generative model.
In our setting, the goal of variational inference is to approximate the true, intractable joint posterior $p(\mathbf{\tilde{y}}_1,\mathbf{\tilde{y}}_2,\mathbf{\tilde{z}}_1,\mathbf{\tilde{z}}_2|\mathbf{x}_1,\mathbf{x}_2)$ with an NN-parameterized variational density $q(\mathbf{\tilde{y}}_1,\mathbf{\tilde{y}}_2,\mathbf{\tilde{z}}_1,\mathbf{\tilde{z}}_2|\mathbf{x}_1,\mathbf{x}_2)$ by minimizing their KL divergence, which is given by the following proposition.

\begin{proposition}
\label{prop 1}
The expectation of the KL divergence between $q(\mathbf{\tilde{y}}_1,\mathbf{\tilde{y}}_2,\mathbf{\tilde{z}}_1,\mathbf{\tilde{z}}_2|\mathbf{x}_1,\mathbf{x}_2)$ and  $p(\mathbf{\tilde{y}}_1,\mathbf{\tilde{y}}_2,\mathbf{\tilde{z}}_1,\mathbf{\tilde{z}}_2|\mathbf{x}_1,\mathbf{x}_2)$ over the data distribution $p_{\mathbf{x}_1,\mathbf{x}_2}$ is given by \eqref{sscc kl}, where \textit{const} denotes a constant.
\end{proposition}

The proof relies on decomposing the KL divergence and leveraging the conditional dependencies of the variables, particularly the Markov chain $\mathbf{\tilde{y}}_1-\mathbf{\tilde{z}}_1-\mathbf{\tilde{z}}_2-\mathbf{\tilde{y}}_2$.
The complete derivation is provided in Appendix A.
Taking a closer look at each term in \eqref{sscc kl}, we can find that the first term corresponds to the distortion of the correlated sources. 
Minimizing this term is equivalent to minimizing the expected distortion of the reconstructed images.
The second and third terms are the code rate required to transmit the latent representations, while the fourth term corresponds to the code rate needed for the hyperpriors.

Based on Proposition 1 and the discussion above, we thus define the loss function of the D-NTSC scheme as
\begin{align}
\mathcal{L}_{\text{SSCC}} &= \mathbb{E}_{p_{\mathbf{x}_1, \mathbf{x}_2}} \Big[ 
    \beta_1 d(\mathbf{x}_1, \mathbf{\hat{x}}_1) + \beta_2 d(\mathbf{x}_2, \mathbf{\hat{x}}_2) \nonumber \\
     - 
    &\log p(\mathbf{\tilde{y}}_1|\mathbf{\tilde{z}}_1) 
    - \log p(\mathbf{\tilde{y}}_2|\mathbf{\tilde{z}}_2) -\log p(\tilde{\mathbf{z}}_1,\mathbf{\tilde{z}}_2) 
\Big],
\label{sscc loss}
\end{align}
where $d(\cdot,\cdot)$ represents the distortion measure between the source $\mathbf{x}_i$ and its recovery $\mathbf{\hat{x}}_i$, and $\beta_1$ and $\beta_2$ control the trade-off between rate and distortion for User 1 and User 2, respectively.

\subsection{Variational Inference for D-NTSCC}

\begin{figure}[t]
    \centering
    \includegraphics[width=0.49\textwidth]{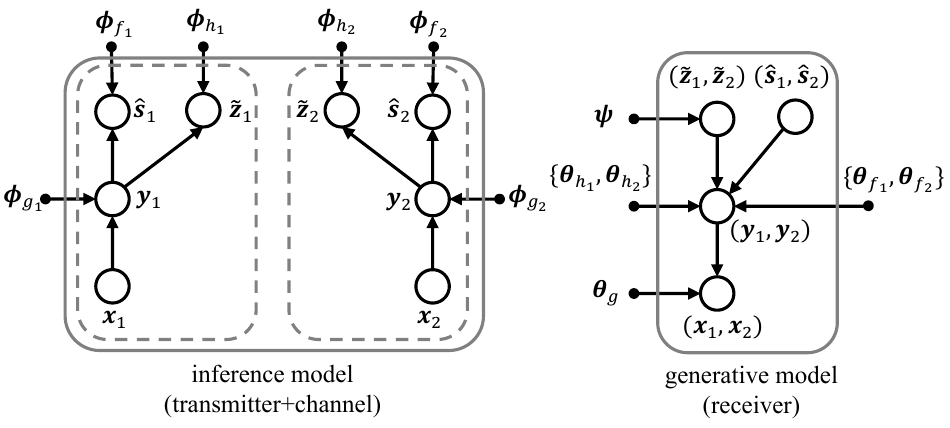}
    \caption{\small{Graphical representation for D-NTSCC.}}
    \label{graphical ntscc}
    \vspace{-0.6cm}
\end{figure}

The VAE model for D-NTSCC is shown in Fig.~\ref{graphical ntscc}.
Similarly, we examine the KL divergence between the NN-parameterized variational density $q(\hat{\mathbf{s}}_1,\hat{\mathbf{s}}_2,\tilde{\mathbf{z}}_1,\tilde{\mathbf{z}}_2|\mathbf{x}_1,\mathbf{x}_2)$ and the intractable joint posterior $p(\hat{\mathbf{s}}_1,\hat{\mathbf{s}}_2,\tilde{\mathbf{z}}_1,\tilde{\mathbf{z}}_2|\mathbf{x}_1,\mathbf{x}_2)$.

\begin{proposition}
\label{prop 2}
The expectation of the KL divergence between $q(\hat{\mathbf{s}}_1,\hat{\mathbf{s}}_2,\tilde{\mathbf{z}}_1,\tilde{\mathbf{z}}_2|\mathbf{x}_1,\mathbf{x}_2)$ and  $p(\hat{\mathbf{s}}_1,\hat{\mathbf{s}}_2,\tilde{\mathbf{z}}_1,\tilde{\mathbf{z}}_2|\mathbf{x}_1,\mathbf{x}_2)$ over the data distribution $p_{\mathbf{x}_1,\mathbf{x}_2}$ is given by \eqref{kl}, where $\text{const}''$ denotes a constant.
\end{proposition}


The derivation follows a similar procedure to the SSCC case and is provided in Appendix~B.  
Specifically, the first term in \eqref{kl} reflects the source distortion.  
In the second and third terms, the probability $p(\mathbf{\hat{s}}_i|\mathbf{\tilde{z}}_i)$ is the convolution of $p(\mathbf{s}_i|\mathbf{\tilde{z}}_i)$ with the Gaussian noise.  
Therefore, the second and third terms can be interpreted as the cost of transmitting $\mathbf{s}_1$ and $\mathbf{s}_2$, respectively, which are proportional to the rates $-\log p(\mathbf{\tilde{y}}_1|\mathbf{\tilde{z}}_1)$ and $-\log p(\mathbf{\tilde{y}}_2|\mathbf{\tilde{z}}_2)$.
The fourth term accounts for the code rate required to transmit the hyperpriors.  

Accordingly, the loss function of the D-NTSCC scheme is defined as
\begin{align}
\mathcal{L}_{\text{JSCC}} &= \mathbb{E}_{p_{\mathbf{x}_1, \mathbf{x}_2}} \Big[ 
    \alpha_1 d(\mathbf{x}_1, \mathbf{\hat{x}}_1) + \alpha_2 d(\mathbf{x}_2, \mathbf{\hat{x}}_2) \nonumber \\
     -\eta 
    &\log p(\mathbf{\tilde{y}}_1|\mathbf{\tilde{z}}_1) 
    - \eta \log p(\mathbf{\tilde{y}}_2|\mathbf{\tilde{z}}_2) -\log p(\tilde{\mathbf{z}}_1,\mathbf{\tilde{z}}_2) 
\Big],
\label{jscc loss}
\end{align}
where $\alpha_1$ and $\alpha_2$ control the trade-off between rate and distortion for User 1 and User 2, respectively, and $\eta$ defines the proportionality between the code rate of the latent representations and the transmission cost of the corresponding channel input.

\textit{Remarks.}
While the SSCC and JSCC loss functions share a similar structure, both consisting of distortion terms and rate terms associated with latent representations and hyperpriors, they differ fundamentally in their designing approach. 
The former follows a modular design, whereas the latter enables joint optimization of the source and channel coding by modeling the channel effects directly within the end-to-end learning process.

\subsection{Joint Probability Function of Hyperprior Variables}

In this subsection, we present the code rate estimation of the hyperpriors $(\mathbf{z}_1, \mathbf{z}_2)$ by defining their joint probability function.  
Note that for univariate probability modeling, Ball\'{e} \textit{et al.}\cite{hyperprior} used a method based on the cumulative distribution function (CDF).
By ensuring that an NN-parameterized function $F(x):\mathbb{R}\rightarrow [0,1]$ satisfies the properties of a valid CDF -- $F(-\infty)=0$, $F(+\infty)=1$, and $\frac{\partial F(x)}{\partial x}\ge 0$ -- this function can be utilized to model the probability distribution of the univariate hyperprior.
However, applying this method to bivariate variables is challenging, since it is difficult to maintain the properties of a valid joint CDF with NNs.

Therefore, we propose a bivariate probability modeling method based on PDFs.
We model $(\mathbf{z}_1,\mathbf{z}_2)$ using a parametric, pairwise factorized density model, where each pairwise joint density $p(z_1^j,z_2^j)$ is parameterized as a Gaussian Mixture Model (GMM). 
Specifically, the density is expressed as
\begin{equation}
\setlength\abovedisplayskip{4pt}
\setlength\belowdisplayskip{2pt}
    p(\mathbf{z}_1,\mathbf{z}_2)=\prod_j p(z_1^j,z_2^j),
\end{equation}
where $p(z_1^j,z_2^j)$ is defined as a mixture of Gaussians
\begin{equation}
\setlength\abovedisplayskip{4pt}
\setlength\belowdisplayskip{5pt}
    p(z_1^j,z_2^j)=\sum_{k=1}^K \pi_k^j \mathcal{N}(\mathbf{m}_k^j,\boldsymbol{\Sigma}_k^j)(z_1^j,z_2^j),
\end{equation}
with $\pi_k^j$, $\mathbf{m}_k^j$, and $\boldsymbol{\Sigma}_k^j$ representing the mixture weights, means, and covariances of the $k$-th Gaussian component of the $j$-th element pair $(z_1^j,z_2^j)$, respectively,  and $K$ representing the number of mixtures.
The parameters $\pi_k^j$, $\mathbf{m}_k^j$, and $\boldsymbol{\Sigma}_k^j$, encapsulated by $\boldsymbol{\psi}$, are learnable parameters optimized by NNs.
Furthermore, the density $p(\mathbf{\tilde{z}}_1,\mathbf{\tilde{z}}_2)$ is obtained by convolving $p(\mathbf{z}_1,\mathbf{z}_2)$ with a bivariate uniform density to approximate the density of the quantized hyperpriors $(\mathbf{\bar{z}}_1,\mathbf{\bar{z}}_2)$, and can be written as
\begin{equation}
    \setlength\abovedisplayskip{4pt}
\setlength\belowdisplayskip{5pt}
p(\mathbf{\tilde{z}}_1,\mathbf{\tilde{z}}_2)=\prod_j\left [\sum_{k=1}^K \pi_k^j \left(\mathcal{N}(\mathbf{m}_k^j,\boldsymbol{\Sigma}_k^j)*\mathcal{U}([-\frac{1}{2}, \frac{1}{2}]^2)\right)(\tilde{z}_1^j,\tilde{z}_2^j)\right],
\end{equation}
where $\mathcal{U}([-\frac{1}{2}, \frac{1}{2}]^2)$ represents the bivariate uniform distribution.
Finally, the probability at integer points is obtained by integrating the density $p(\mathbf{\tilde{z}}_1,\mathbf{\tilde{z}}_2)$ over corresponding quantization intervals (which is differentiable).
Note that this method can be extended to multiple users by generalizing the bivariate Gaussian variables to multivariate Gaussian variables.

\textit{Remarks.}
Joint density modeling of the hyperpriors, as opposed to assuming their independence, offers several advantages.
First, it enables a more accurate entropy model for estimating the true joint distribution $p(\mathbf{\tilde{y}}_1, \mathbf{\tilde{y}}_2,\mathbf{\tilde{z}}_1,\mathbf{\tilde{z}}_2)$, thereby improving rate-distortion performance.
Specifically, the joint entropy model is defined as $p(\mathbf{\tilde{y}}_1, \mathbf{\tilde{y}}_2,\mathbf{\tilde{z}}_1,\mathbf{\tilde{z}}_2)=p(\mathbf{\tilde{z}}_1,\mathbf{\tilde{z}}_2)p(\mathbf{\tilde{y}}_1|\mathbf{\tilde{z}}_1)p(\mathbf{\tilde{y}}_2|\mathbf{\tilde{z}}_2)$, instead of assuming independent hyperpriors $p(\mathbf{\tilde{z}}_1)p(\mathbf{\tilde{z}}_2)p(\mathbf{\tilde{y}}_1|\mathbf{\tilde{z}}_1)p(\mathbf{\tilde{y}}_2|\mathbf{\tilde{z}}_2)$ as in\cite{li2024content}, which does not account for source correlations.
Second, since both transmitters share this joint probability function, they can predict each other's transmission rate and adjust encoding accordingly.
Specifically, given $\mathbf{z}_1$ and the joint probability function, User 1 can estimate $\mathbf{z}_2$ using the MMSE estimator, denoted as $\mathbf{z}_2^\ast$. 
By applying $h_{s_2}(\cdot)$, User 1 can obtain an estimate of $\boldsymbol{\sigma}_2$ and $\boldsymbol{\mu}_2$, thus an estimate of the code rate of $\mathbf{y}_2$, which we denote as $-\log p(\mathbf{y}_2|\mathbf{z}_2^\ast)$. 
This estimated code rate is then incorporated in the the JSCC encoder $f_{e_1}$, facilitating the encoding of $\mathbf{y}_1$.

\section{Neural Network Architecture Design}

This section presents the detailed NN architectures, including the nonlinear transform modules, the transformation module, and the JSCC codecs.
The nonlinear transform modules and the transformation module are present in both D-NTSC and D-NTSCC, and share the same architecture, whereas the JSCC codecs are used only in D-NTSCC.

\subsection{Nonlinear Transform Modules}

\begin{figure}[t]
    \centering
    \begin{subfigure}[b]{0.42\textwidth}
        \centering
        \includegraphics[width=\textwidth]{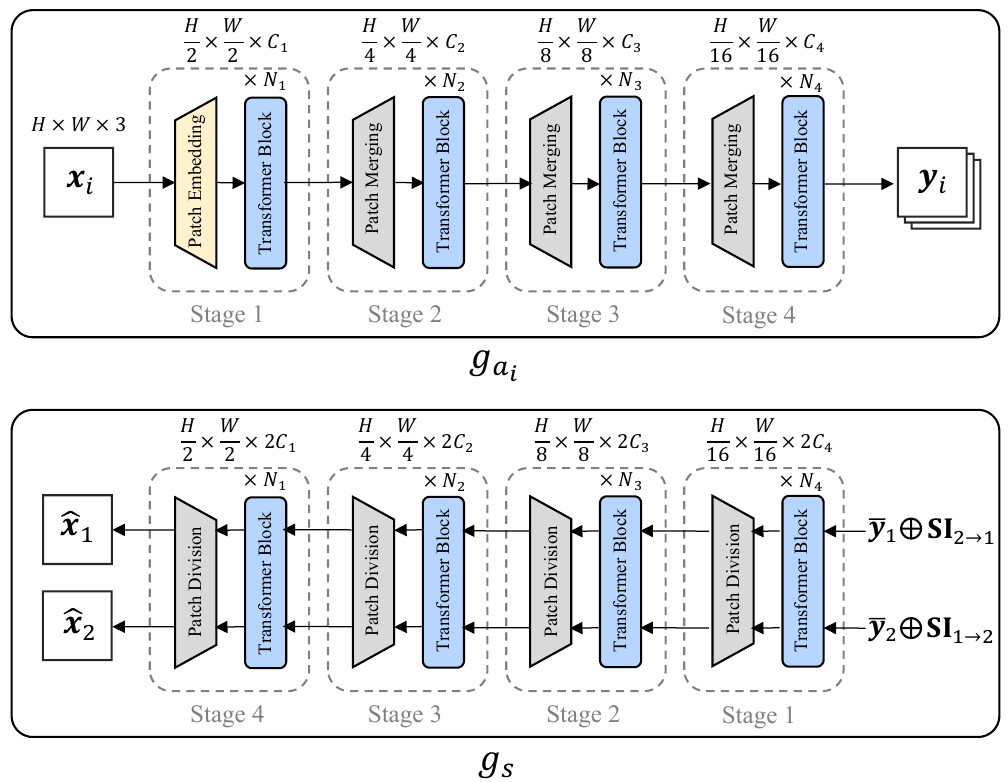}
        \caption{Nonlinear transforms $g_{a_i}$ and $g_s$.}
    \end{subfigure}
    \hfill
    \begin{subfigure}[b]{0.42\textwidth}
        \centering
        \includegraphics[width=\textwidth]{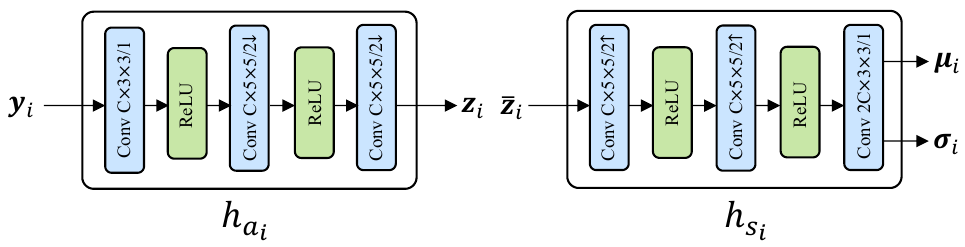}
        \caption{Analysis transform $h_{a_i}$ and synthesis transform $h_{s_i}$.}
    \end{subfigure}
    \caption{NN architectures of nonlinear transform modules.}
    \label{ntc modules}
    \vspace{-0.6cm}
\end{figure}

\begin{figure}[t]
    \centering
    \includegraphics[width=0.45\textwidth]{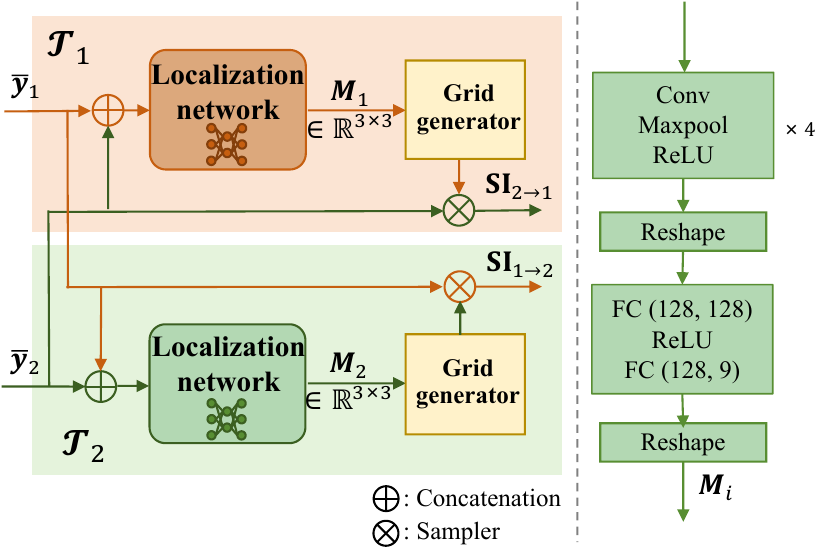}
    \caption{Left: Transformation module. Right: Detailed architecture of the localization network.}
    \label{stn}
    \vspace{-0.6cm}
\end{figure}

\begin{figure*}[t]
    \centering
    \begin{subfigure}[b]{0.43\textwidth}
        \centering
        \includegraphics[width=\textwidth]{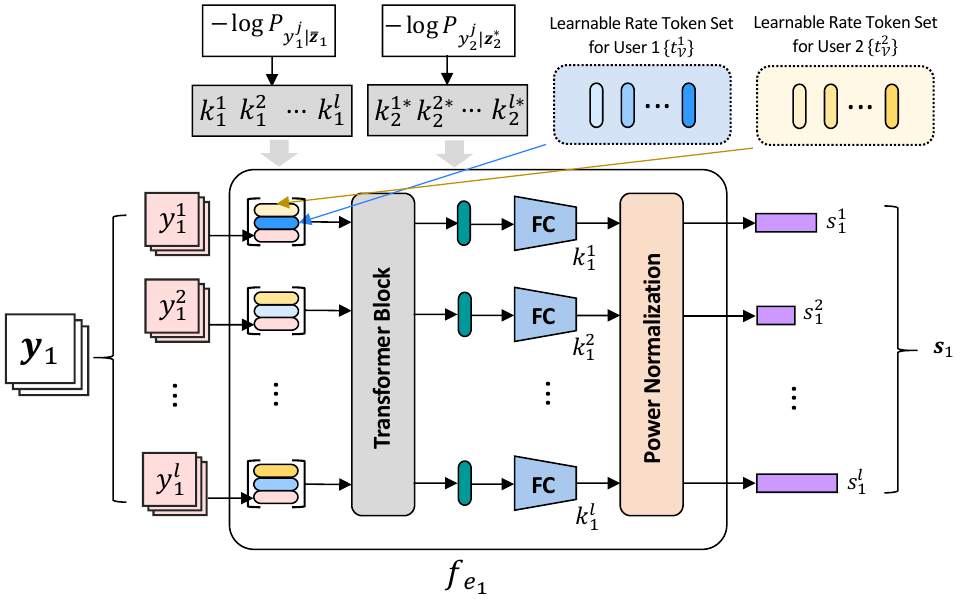}
        \caption{JSCC encoder $f_{e_1}$.}
    \end{subfigure}
    \begin{subfigure}[b]{0.43\textwidth}
        \centering
        \includegraphics[width=0.9\textwidth]{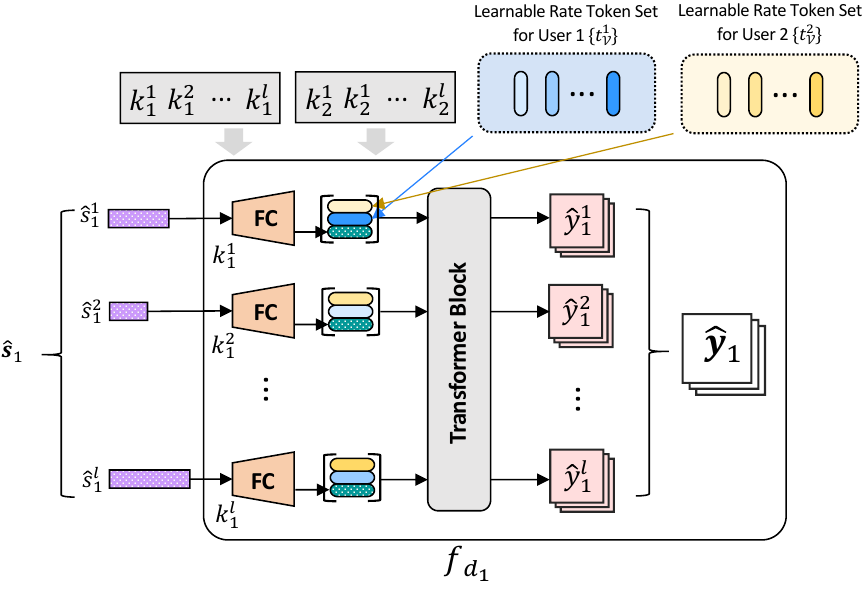}
        \caption{JSCC decoder $f_{d_1}$.}
    \end{subfigure}
    \caption{NN architectures of JSCC codecs.}
    \label{jscc modules}
    \vspace{-0.7cm}
\end{figure*}

We adopt the Swin Transformer architecture \cite{liu2021swin} as the backbone of the nonlinear transforms.
Fig.~\ref{ntc modules}(a) shows the NN architectures of the nonlinear transforms $g_{a_i}$ and $g_s$.
The analysis transform $g_{a_i}$ and the synthesis transform $g_s$ follow a symmetric structure.
In $g_{a_i}$, each image passes through four stages of transformer blocks, gradually downsampled into a latent representation $\mathbf{y}_i\in\mathbb{R}^{\frac{H}{16}\times\frac{W}{16}\times C_4}$.
Conversely, the function $g_s$ upsamples the concatenated latent representations $\mathbf{\bar{y}}_1\oplus\mathbf{SI}_{2\rightarrow 1}$ and $\mathbf{\bar{y}}_2\oplus\mathbf{SI}_{1\rightarrow 2}$ (or $\mathbf{\hat{y}}_1\oplus\mathbf{SI}_{2\rightarrow 1}$ and $\mathbf{\hat{y}}_2\oplus\mathbf{SI}_{1\rightarrow 2}$ in D-NTSCC) into $\mathbf{\hat{x}}_1$ and $\mathbf{\hat{x}}_2$, respectively, where $\oplus$ denotes the concatenation operation.
Additionally, the NN architectures for the analysis and synthesis transforms of the hyperprior $\mathbf{z}_i$ are summarized in Fig.~\ref{ntc modules}(b).
These architectures consist of convolutional layers combined with ReLU activation functions.

\subsection{Transformation Module}

Since the images are obtained at physically separated locations, they may exhibit perspective differences, which will also be reflected in the extracted latent representations. 
This misalignment makes it difficult for $g_s$ to effectively utilize the other latent representation as side information.
Therefore, we propose a transformation module designed to align the two recovered latent representations prior to the synthesis transform, thereby enabling $g_s$ to better exploit their correlation.
We seek to achieve this goal by using spatial transformers (STs)\cite{stn}, which spatially transform latent representations in a learnable and differentiable manner.
For each input, STs produce a transformation matrix via a localization network.
The transformation is then applied to the input by a grid generator and a sampler.
This allows STs to adjust regions of interest in a latent representation to an expected pose, facilitating easy processing in the following layers.

We relate the two images using a projective transformation, which models the geometric relationship between them as a homography.
Given a pixel coordinate in one image $\mathbf{u}=(w,h,1)$, its corresponding coordinate in the other image $\mathbf{u}'=(w',h',d')$ is obtained via a projective transformation matrix $\mathbf{M}\in\mathbb{R}^{3\times3}$
\begin{equation}
    \mathbf{u}'=\mathbf{M}\mathbf{u}=
\begin{bmatrix}
m_{11} & m_{12} & m_{13} \\
m_{21} & m_{22} & m_{23} \\
m_{31} & m_{32} & m_{33}
\end{bmatrix}
\begin{bmatrix}
w \\ h \\ 1
\end{bmatrix}.
\label{trans}
\end{equation}
The final Cartesian coordinates in the other image $\mathbf{u}''=(w'',h'')$ are obtained by normalization: $w''=\frac{w'}{d'}$, $h''=\frac{h'}{d'}$.
The perspective transformation can relate any two images of the same planar surface under the assumption of a pinhole camera model\cite{perspective}, which makes it particularly useful for aligning images captured from different viewpoints.

Fig.~\ref{stn} illustrates the architecture of the transformation module.
Specifically, $\mathcal{T}_1$ and $\mathcal{T}_2$ take concatenated inputs of $\mathbf{\bar{y}}_1$ and $\mathbf{\bar{y}}_2$ (or $\mathbf{\hat{y}}_1$ and $\mathbf{\hat{y}}_2$ in D-NTSCC).
These concatenated inputs are first passed through a localization network, consisting of convolutional layers and fully connected (FC) layers, as shown on the right side of Fig.~\ref{stn}, to produce a pointwise projective transformation matrix $\mathbf{M}_i$ ($i \in \{1, 2\}$).
Here, \textit{pointwise} indicates that the same projective transformation is applied to each element (each ``pixel'') of the received latent representation.
The matrices $\mathbf{M}_1$ and $\mathbf{M}_2$ are then used by the grid generator to create sampling grids for $\mathbf{\bar{y}}_2$ and $\mathbf{\bar {y}}_1$, respectively, specifying sampling locations to produce $\mathbf{SI}_{2\to 1}$ and $\mathbf{SI}_{1\to 2}$, as in \eqref{trans}.
A sampler applies the transformation by sampling the corresponding $\mathbf{\bar {y}}_i$ (or $\mathbf{\hat {y}}_i$ in D-NTSCC) at the specified grid points, using methods such as bilinear interpolation\cite{stn}.
This whole process is differentiable, allowing for the parameter updates of the localization network through standard stochastic gradient descent (SGD) algorithms.

\subsection{JSCC Codecs}

In D-NTSCC, the JSCC encoder $f_{e_i}$ at the transmitter further compresses $\mathbf{y}_i$ according to its code rate, generating the channel input symbols $\mathbf{s}_i$ with adaptive lengths, as illustrated in Fig. \ref{jscc modules}(a).
Note that the JSCC codecs are identical across users; in what follows, we use $f_{e_1}$ and $f_{d_1}$ as representatives for explanation.
Specifically, for User 1, the latent representation $\mathbf{y}_1$ is first reshaped to the size of $l\times C_4$, where $l=\frac{H}{16}\cdot\frac{W}{16}$.
Thus, $\mathbf{y}_1$ can be written as $\mathbf{y}_1=(y_1^1, y_1^2, ..., y_1^l)$, where each element has a length of $C_4$.
The entropy sum of $y_1^j$, for $j=1,...,l$, denoted as $-\log P_{y_1^j|\mathbf{\bar{z}}_1}$, is used to determine the bandwidth cost of $s_1^j$, denoted as $k_1^j$.
The value $k_1^j$ belongs to a set of integers $\mathcal{V}=\{v_1, v_2, ..., v_q\}$, and is determined by
\begin{equation}
\setlength\abovedisplayskip{4pt}
\setlength\belowdisplayskip{5pt}
    k_1^j = \underset{v\in\mathcal{V}}{\mathrm{argmin}}\left|-\eta \log P_{y_1^j|\mathbf{\bar{z}}_1} - v \right|.
\end{equation}
The JSCC encoder in the proposed D-NTSCC extends the one in \cite{ntscc} by additionally incorporating the statistics of the other user.
The estimated entropy sum of $y_2^j$, denoted as $-\log P_{y_2^j|\mathbf{z}_2^\ast}$, is utilized to compute the estimated bandwidth cost of $s_2^j$, denoted as $k_2^{j\ast}$.
That is, the encoder adapts its output based not only on its own rate information but also on the rate statistics of the other user. 
To indicate the rate information, two sets of learnable rate tokens, $\{t_\mathcal{V}^1\}$ and $\{t_\mathcal{V}^2\}$, are introduced for User 1 and User 2, respectively, where each token corresponds to a specific element in the rate set $\mathcal{V}$.
The corresponding tokens are selected and concatenated with $y_1^j$, which is then passed into a Transformer block, leveraging its self-attention mechanism to effectively integrate the rate information.
Then, a set of variable-length FC layers encodes $y_1^j$ into the desired dimensions, resulting in $\mathbf{s}_1$ with variable-length elements.

At the receiver, the JSCC decoder maps $\mathbf{\hat{s}}_1$ to the recovered latent representation $\mathbf{\hat{y}}_1\in\mathbb{R}^{l\times C_4}$.
This process serves as the reverse operation of the JSCC encoder, as illustrated in Fig.~\ref{jscc modules}(b), where each element of $\mathbf{\hat{s}}_1$ is first processed by an FC layer, then concatenated with the corresponding rate tokens, before being passed through a Transformer block to reconstruct $\mathbf{\hat{y}}_1$.

\begin{figure*}
     \centering
     \begin{subfigure}[b]{0.48\textwidth}
         \centering
         \includegraphics[scale=0.21]{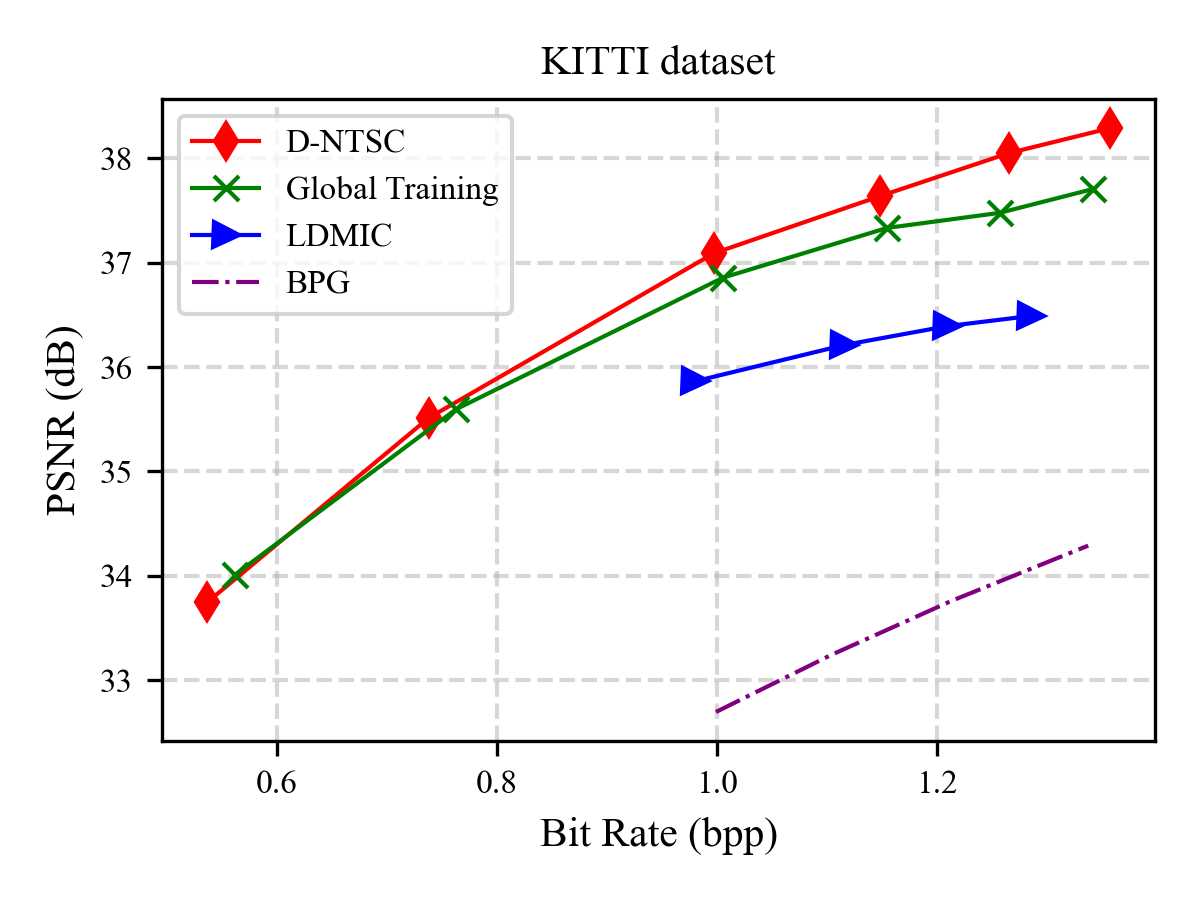}
         \caption{PSNR performances on the \textit{KITTI Stereo} dataset.}
     \end{subfigure}
     \hfill
     \begin{subfigure}[b]{0.48\textwidth}
         \centering
         \includegraphics[scale=0.21]{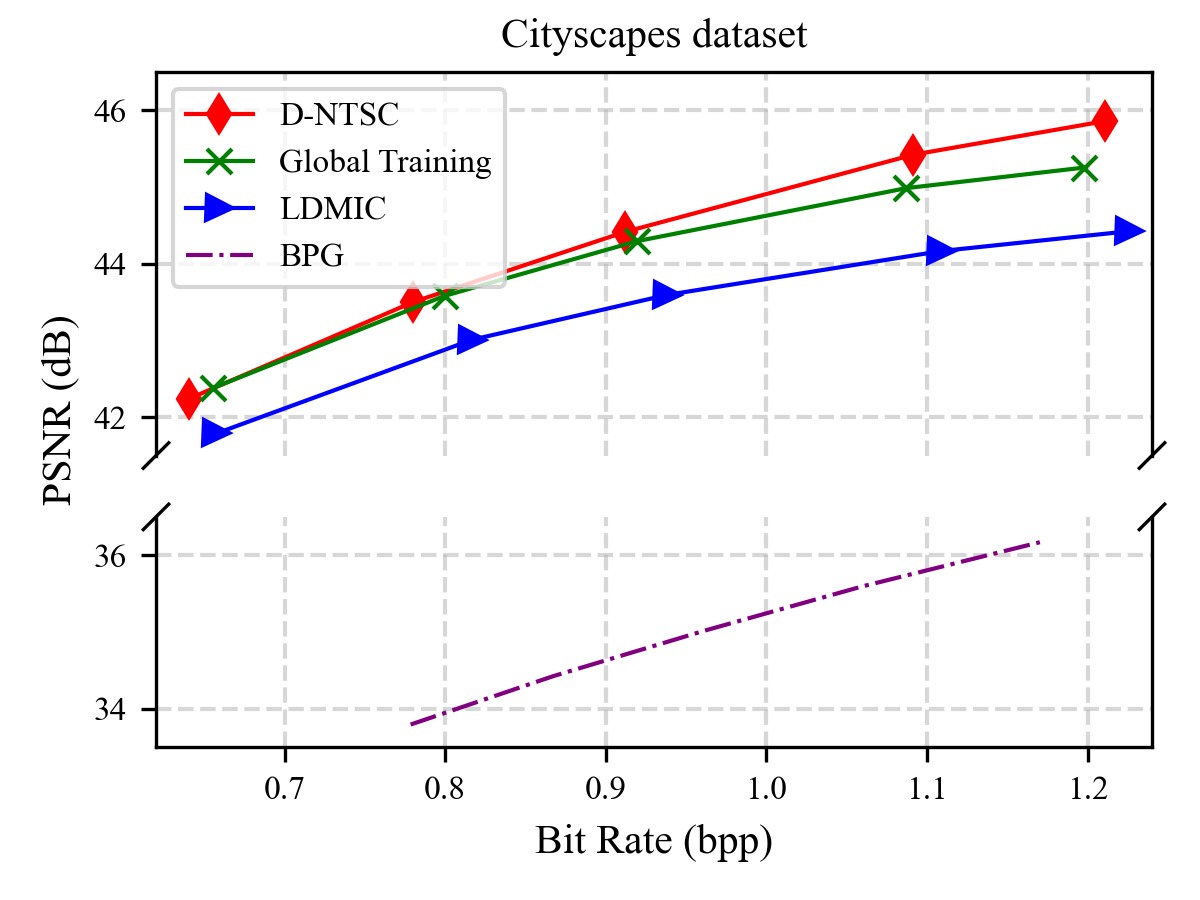}
         \caption{PSNR performances on the \textit{Cityscapes} dataset.}
     \end{subfigure}
     \hfill
     \caption{PSNR performances at varying bit rates.}
     \label{bitrates}
     \vspace{-0.5cm}
\end{figure*}


\section{Experimental Results}

In this section, we validate the performance of the proposed D-NTSC and D-NTSCC schemes through extensive simulations.
Section VI-A outlines the experimental settings.
Section VI-B evaluates D-NTSC by comparing it with existing distributed source coding baselines.
Section VI-C evaluates D-NTSCC across various channel conditions and transmission rates.
In Section VI-D, we compare D-NTSCC and D-NTSC to provide insights into their respective strengths.
An ablation study on the transformation module is presented in Section VI-E to demonstrate its effectiveness.
Finally, Section VI-F compares the computational complexity of the proposed and baseline methods.

\subsection{Experiment settings}

\subsubsection{Dataset}

Two common stereo image datasets are selected to evaluate the performance of the proposed framework.
One is \textit{KITTI Stereo}, which consists of 1578 training stereo image pairs, and 790 testing stereo image pairs.
A stereo image pair refers to a pair of images each taken at the same time from a different camera.
The other is \textit{Cityscapes}, which contains 2975 training stereo image pairs, 500 validation pairs, and 1525 testing pairs.
For preprocessing, each image in the KITTI Stereo dataset is center-cropped to a size of $370\times 740$, and then downsampled to $128\times 256$.
For Cityscape dataset, we directly downsample each image to the size of $128\times 256$.

\subsubsection{Training details and hyperparameters}
The Adam optimizer is employed for training with a batch size of 2.
We use a cosine annealing schedule with an initial learning rate (LR) of $1\times10^{-4}$, which gradually decreases to $1\times10^{-6}$ according to $lr(t)=10^{-6}+\frac{1}{2}(10^{-4}-10^{-6})(1+\cos\frac{t}{N}\pi)$, where $t$ represents the epoch index, and $N$ is the total number of training epochs, which is set to 300 for D-NTSC, and 200 for D-NTSCC.
For the nonlinear transforms, the number of channels is configured as $C_1=128$, $C_2=160$, $C_3=192$, $C_4=256$, while the number of blocks at each stage is $N_1=2$, $N_2=2$, $N_3=6$, $N_4=2$.
The number of attention heads for each stage is set to $[4, 8, 8, 8]$, and the patch size is fixed at 2. 
The rate token length is set to 4.
The set of bandwidth costs of $\mathbf{s}_i$ is $\mathcal{V} = \{8a \mid a = 1, 2, \dots, 20\}$.
The compression ratio of the proposed D-NTSC and D-NTSCC can be controlled through changing the hyperparameters $\beta_i$ in \eqref{sscc loss} and $\alpha_i$ in \eqref{jscc loss}, respectively $(i=1,2)$.
Unless stated otherwise, we set $\beta_1=\beta_2$ and $\alpha_1=\alpha_2$.
The hyperparameter $\eta$ in \eqref{jscc loss} is fixed at 1.
For the number of mixtures, we set $K=1$.

\subsubsection{Benchmarks}
We compare the proposed D-NTSC scheme with the following source coding benchmarks. 

\begin{itemize}

    \item \emph{Global Training}: 
    This baseline serves as an ablation study to evaluate the effectiveness of the proposed method in leveraging source correlation. 
    The encoder and decoder adopt the same Swin Transformer backbone but are trained globally on all available images, regardless of camera view. 
    The trained model is then applied independently to both users for local encoding and decoding.
    
    \item \emph{LDMIC\cite{ldmic}}: 
    A state-of-the-art distributed multi-view image coding method that introduces a joint context transfer module at the decoder to exploit global inter-view correlations.
    
    \item \emph{BPG\cite{bellard2014bpg}}:
    The Better Portable Graphics (BPG) codec, a widely used standard for efficient image compression, serving here as a traditional source coding baseline.
    
\end{itemize}

We compare the proposed D-NTSCC scheme with the following DL-based JSCC benchmarks, as well as a conventional transmission scheme using BPG and capacity-achieving channel codes. 

\begin{itemize}

    \item \emph{D-DJSCC \cite{wang2022distributed}}: 
    A distributed deep JSCC scheme that exploits source correlation via a CA module.
    
    \item \emph{DHF-JSCC \cite{li2024content}}: 
    A recent distributed deep JSCC scheme that introduces a multi-layer CA-based information fusion module, allowing pixel-level correlation exploitation across views.
    
    \item \emph{NTSCC \cite{ntscc}}:
    The NTSCC scheme designed for point-to-point communication.
    Like \emph{Global Training}, the model is trained on all available data without considering camera view and is applied to both users.
    
    \item \emph{BPG+Capacity}:
    A conventional SSCC scheme that assumes a capacity-achieving channel code and does not consider source correlation.
\end{itemize}


\subsubsection{Metrics}
Two performance metrics are considered. 
One is the peak signal-to-noise ratio (PSNR), where the distortion measure in the loss function \eqref{sscc loss} and \eqref{jscc loss} is set to be the mean-square-error (MSE). 
The other is the multi-scale structural similarity index (MS-SSIM), where the distortion measure is set to be $1-\text{MS-SSIM}$. 


\begin{figure}[t]
    \centering
    \includegraphics[width=0.48\textwidth]{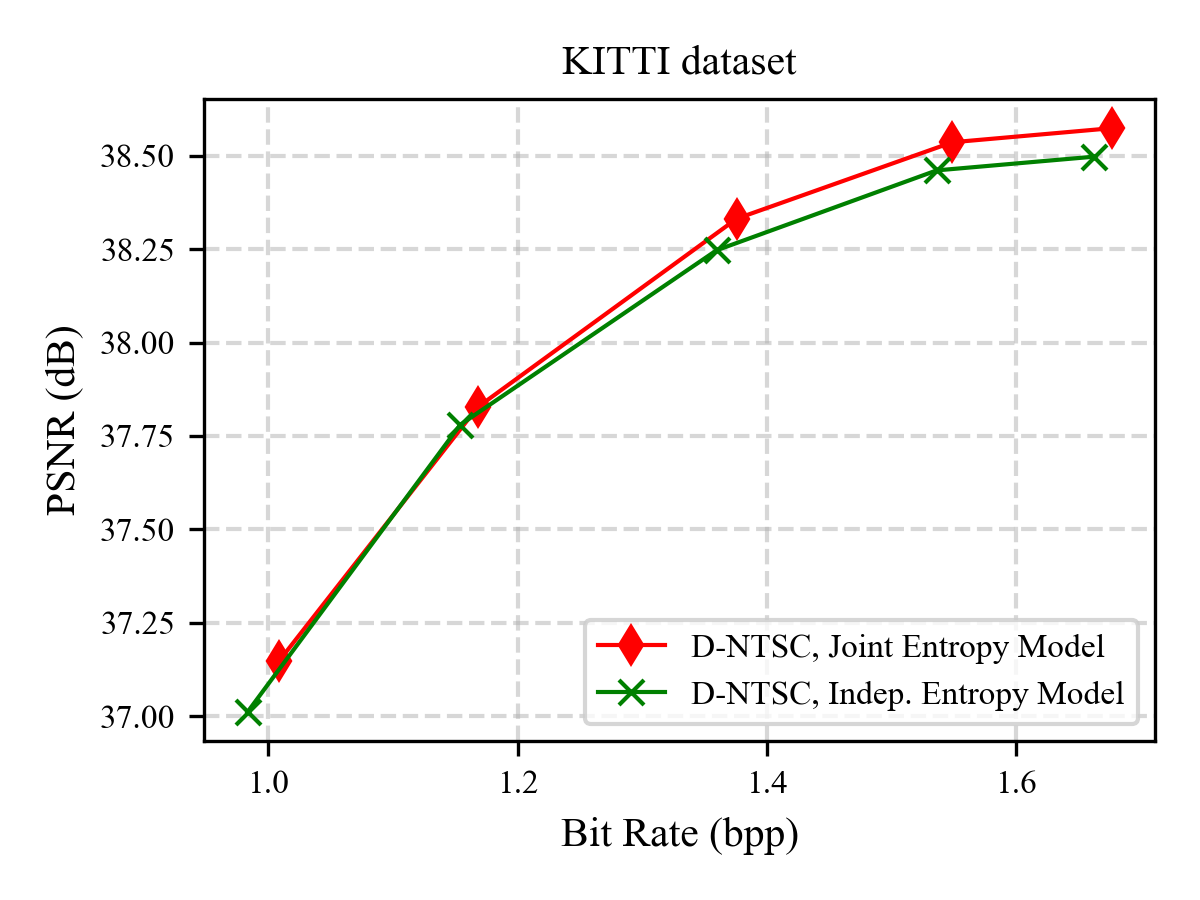}
    \caption{Ablation study on the joint entropy model.}
    \label{ablation prob}
\end{figure}

\begin{figure}
     \centering
     \begin{subfigure}[b]{0.48\textwidth}
         \centering
         \includegraphics[scale=0.21]{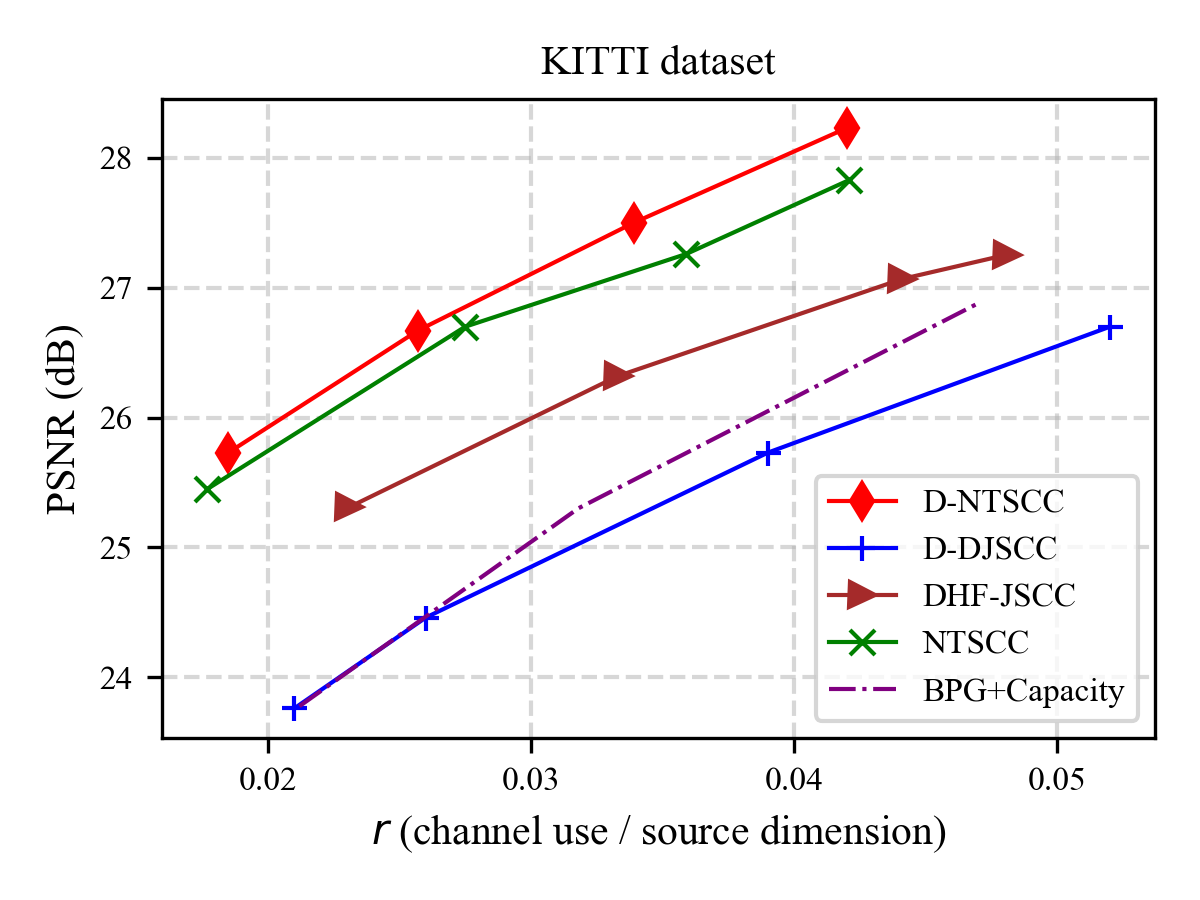}
         \caption{PSNR performances.}
     \end{subfigure}
     \hfill
     \begin{subfigure}[b]{0.48\textwidth}
         \centering
         \includegraphics[scale=0.21]{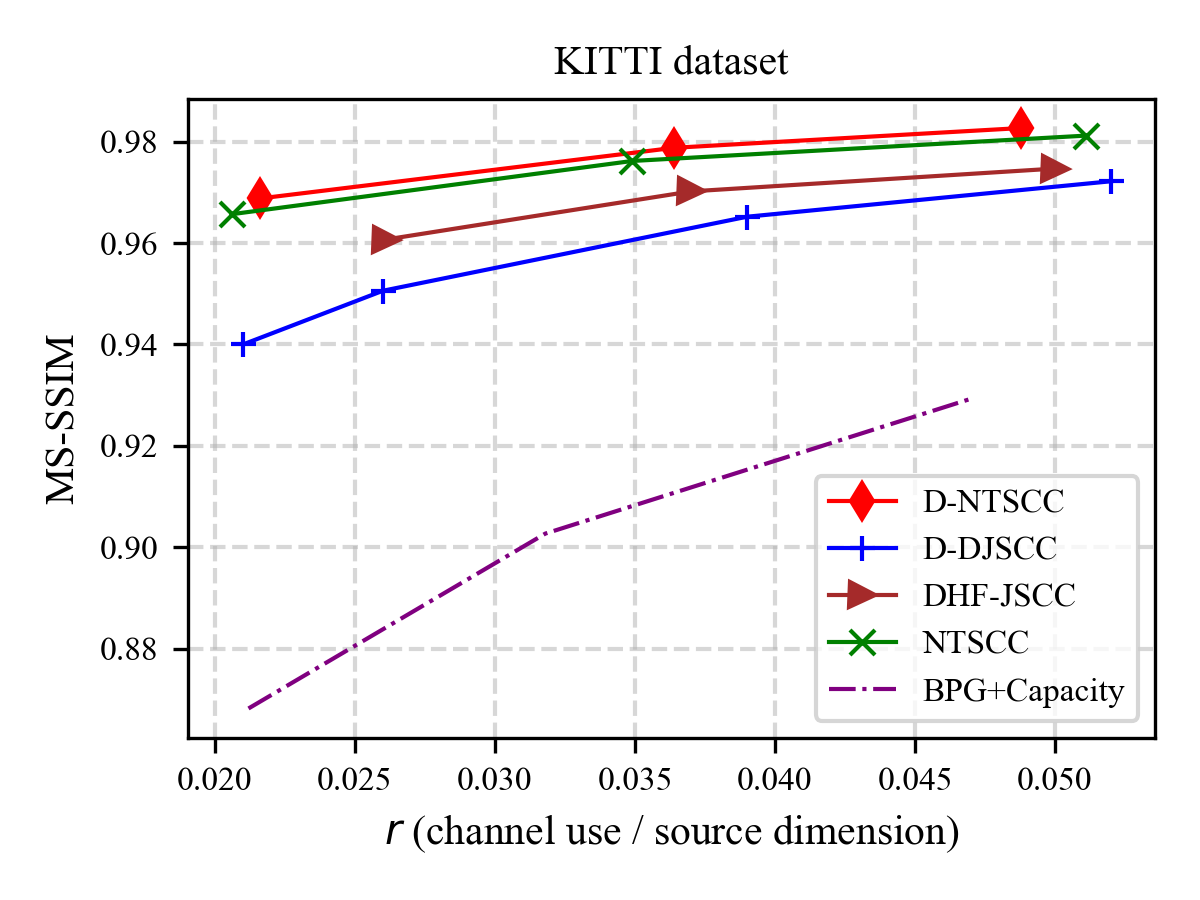}
         \caption{MS-SSIM performances.}
     \end{subfigure}
     \hfill
        \caption{Performances at varying transmission rates when channel SNR = 5 dB.}
        \label{transmisstion rate}
\end{figure}

\subsection{Evaluation of D-NTSC}

We first compare the proposed D-NTSC with distributed source coding baselines. 
We report performance in terms of PSNR versus bit rate, where bit rate is defined as the average number of bits used per pixel, expressed as
\begin{equation}
    \textit{bit rate}_{i} = \frac{R_i}{HW}\ \ \ \ bit/pixel,
\end{equation}
where $R_i$ is the total code rate defined in \eqref{code rate}.

Fig.~\ref{bitrates}(a) and Fig.~\ref{bitrates}(b) plot PSNR as a function of bit rate, for the KITTI Stereo and the Cityscapes datasets, respectively.
Note that since the performance of the two users is nearly identical for all schemes, we report the results for User 1 without loss of generality.
For both datasets, the proposed scheme achieves the highest performance across a wide range of bit rates.
Specifically, D-NTSC outperforms the global training scheme by 0.3 dB for the KITTI Stereo dataset at a bit rate of 1.0 bpp.
This gap widens to 0.6 dB at 1.35 bpp.
A similar trend can be observed on the Cityscapes dataset.
These gains stem from the effective use of side information in the proposed scheme.
Furthermore, compared with the distributed coding scheme, LDMIC, the proposed method achieves a bit rate reduction of 30.7$\%$ on the KITTI Stereo dataset and 26.0$\%$ on the Cityscapes dataset.
Besides the probabilistic and geometric correlation modeling, this performance gain also results from architectural differences.
While LDMIC relies on CNNs combined with a CA-based module, the proposed scheme is built upon the Swin Transformer, which enables more effective feature extraction and correlation learning at both local and global levels through its hierarchical structure and shifted window mechanism.
Additionally, the proposed scheme outperforms BPG significantly on both datasets.

Furthermore, to evaluate the impact of the proposed joint entropy model, we conduct an ablation study comparing its performance against that of using an independent entropy model, $p(\mathbf{\tilde{z}}_1)p(\mathbf{\tilde{z}}_2)p(\mathbf{\tilde{y}}_1|\mathbf{\tilde{z}}_1)p(\mathbf{\tilde{y}}_2|\mathbf{\tilde{z}}_2)$.
The results are illustrated in Fig.~\ref{ablation prob}.
As can be seen, the joint entropy model improves the rate-distortion performance, especially at higher bit rates, reducing the bit rate by up to 8.43$\%$ compared to the independent entropy model.
This phenomenon can be attributed to its ability to better approximate the true joint distribution of the sources.
These results demonstrate the benefits of explicitly modeling source correlations in distributed compression settings.

\subsection{Evaluation of D-NTSCC}

In this subsection, we evaluate the proposed D-NTSCC by comparing it with existing baselines under varying transmission rates and channel SNR conditions.

\subsubsection{Performance vs. transmission rate}
We first compare the rate-distortion performance of different schemes, where the transmission rate $r$ is defined in \eqref{def r}.
During training, the channel SNR is fixed at 5 dB for both users.

Fig.~\ref{transmisstion rate}(a) illustrates the PSNR performance of different schemes at varying transmission rates.
The results show that the proposed D-NTSCC scheme outperforms all benchmarks across the evaluated transmission rates.
Specifically, compared to the two existing distributed coding schemes, D-DJSCC and DHF-JSCC,  D-NTSCC achieves up to 50$\%$ and 30$\%$ bandwidth savings, respectively.
This performance gain is attributed to the joint entropy modeling and the ``upgrade'' of the NN architecture from CNN to Swin Transformer.
Additionally, the proposed scheme achieves around 10$\%$ bandwidth savings compared to the point-to-point NTSCC scheme, demonstrating its effectiveness in exploiting the source correlation.

Similar trends can be seen in the MS-SSIM curves shown in Fig.~\ref{transmisstion rate}(b).
The D-NTSCC scheme achieves the highest MS-SSIM performance across all evaluated transmission rates, resulting in bandwidth savings of 16$\%$, 50$\%$, and 38$\%$ compared to NTSCC, D-DJSCC, and DHF-JSCC, respectively.

\subsubsection{Performance vs. SNR}

\begin{figure}[t]
    \centering
    \includegraphics[width=0.48\textwidth]{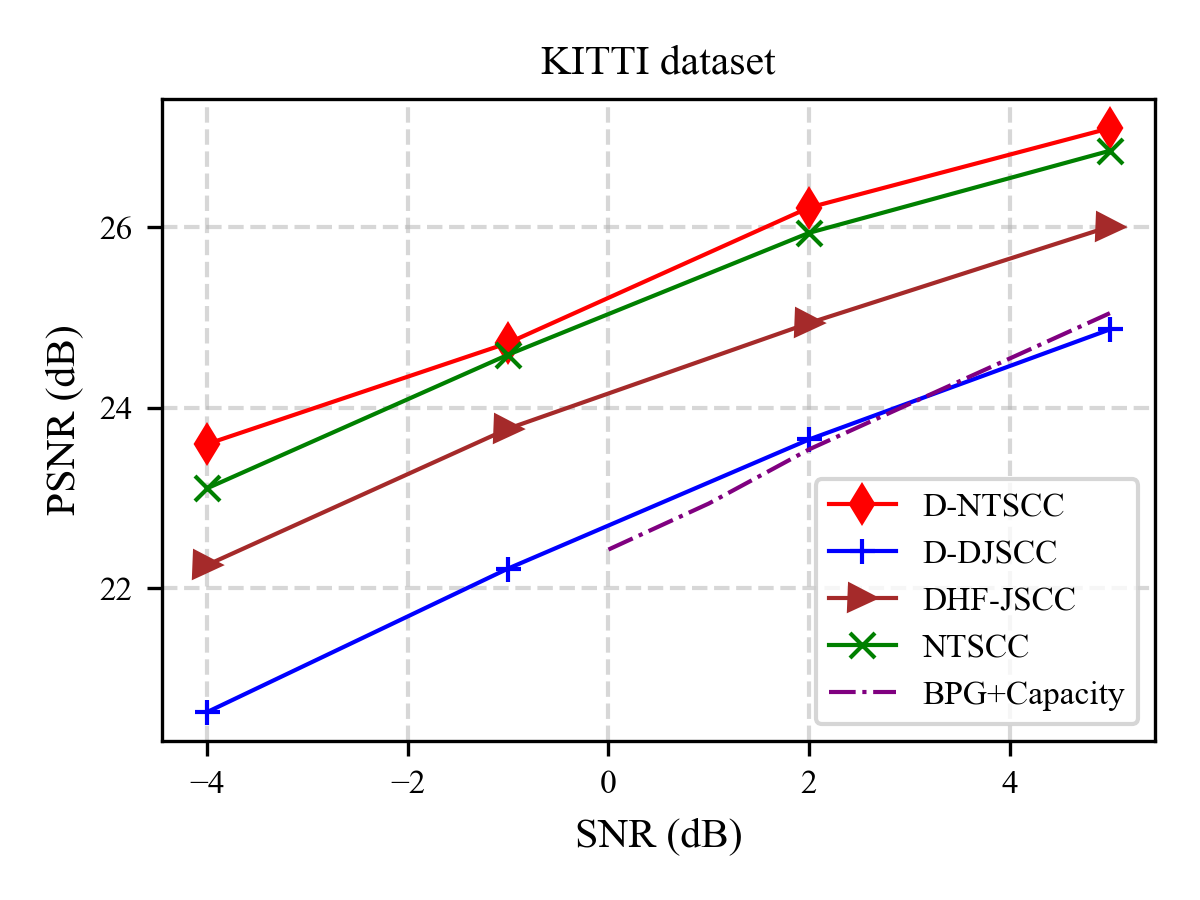}
    \caption{Performances at varying SNRs when $r$ = 0.03.}
    \label{snr}
    \vspace{-0.5cm}
\end{figure}

We further evaluate the PSNR performance of each scheme under varying SNR conditions, with the transmission rate fixed at $r=0.03$. 
The testing SNR is the same as the training SNR.
The results are shown in Fig.~\ref{snr}.
As can be seen, the proposed D-NTSCC achieves the highest PSNR across all evaluated SNRs.
Specifically, it outperforms NTSCC by approximately 0.3 dB, DHF-JSCC by 1 dB, and D-DJSCC and BPG+Capacity by more than 2 dB.
These results demonstrate that the proposed scheme is able to maintain advantages across diverse channel conditions.

\subsection{D-NTSCC vs. D-NTSC+Capacity}

\begin{figure}[t]
    \centering
    \includegraphics[width=0.48\textwidth]{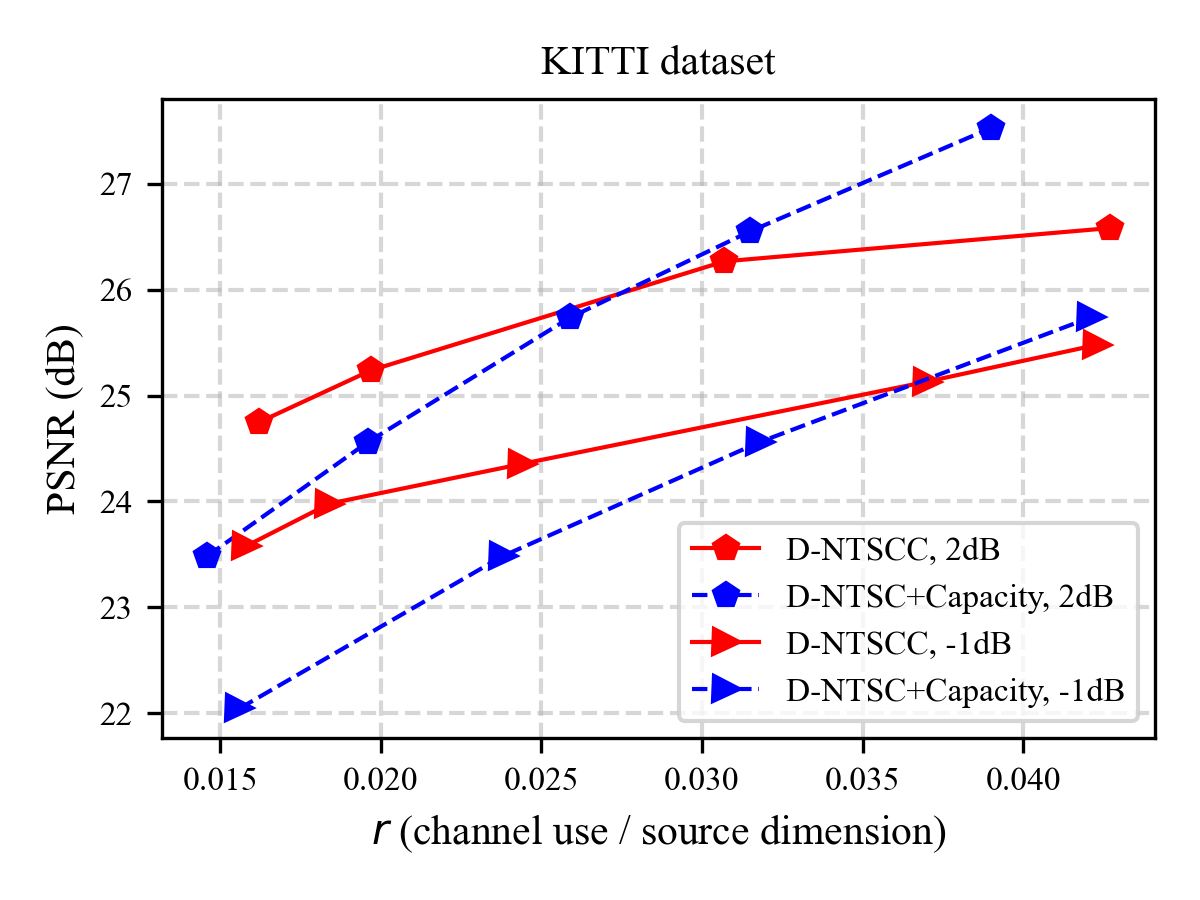}
    \caption{Performance comparison between D-NTSCC and D-NTSC+Capacity.}
    \label{comp proposed}
\end{figure}

Fig.~\ref{comp proposed} compares the rate-distortion performance of the proposed D-NTSCC scheme with the D-NTSC+Capacity scheme under two channel SNRs, 2 dB and -1 dB.
The results show that the JSCC approach achieves better performance at lower rates and lower SNRs.
For example, when $r = 0.02$, D-NTSCC yields a 0.7 dB gain over D-NTSC+Capacity at SNR = 2 dB, and a 1.3 dB gain at SNR = -1 dB.
In contrast, at SNR = 2 dB, D-NTSC+Capacity outperforms D-NTSCC for rates greater than 0.027, and at SNR = -1 dB, D-NTSC+Capacity outperforms D-NTSCC for rates above 0.036.

These observations offer guidance for choosing the appropriate scheme in different scenarios.
D-NTSCC is better suited for resource-constrained environments and challenging, dynamic channels, while D-NTSC is preferable in scenarios where compatibility with existing systems is needed and resources are more abundant.

\subsection{Ablation Study on the Transformation Module}

\begin{figure}[t]
    \centering
    \includegraphics[width=0.48\textwidth]{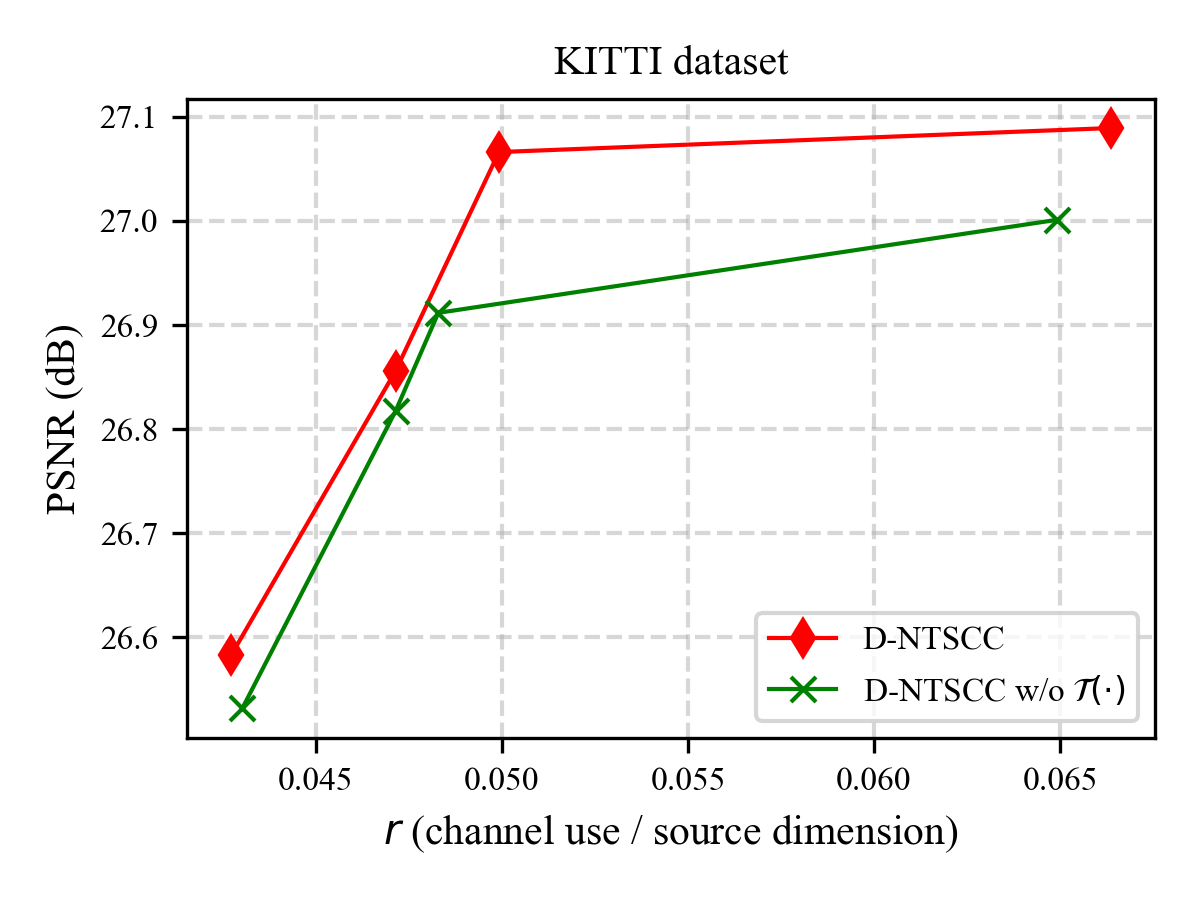}
    \caption{Ablation study on the transformation module.}
    \label{ablation t}
    \vspace{-0.5cm}
\end{figure}

\begin{figure*}
    \begin{small}
    \begin{align}
        &\ \ \ \ \ \mathbb{E}_{p_{\mathbf{x}_1,\mathbf{x}_2}}D_{KL}[q(\mathbf{\tilde{y}}_1,\mathbf{\tilde{y}}_2,\mathbf{\tilde{z}}_1,\mathbf{\tilde{z}}_2|\mathbf{x}_1,\mathbf{x}_2)||p(\mathbf{\tilde{y}}_1,\mathbf{\tilde{y}}_2,\mathbf{\tilde{z}}_1,\mathbf{\tilde{z}}_2|\mathbf{x}_1,\mathbf{x}_2)]\nonumber\\
        &\overset{(a)}{=}\!\mathbb{E}_{p_{\mathbf{x}_1,\mathbf{x}_2}}\mathbb{E}_{q_{\mathbf{\tilde{y}}_1,\mathbf{\tilde{y}}_2,\mathbf{\tilde{z}}_1,\mathbf{\tilde{z}}_2|\mathbf{x}_1,\mathbf{x}_2}}\! \left\{ \log q(\mathbf{\tilde{y}}_1,\mathbf{\tilde{y}}_2,\mathbf{\tilde{z}}_1,\mathbf{\tilde{z}}_2|\mathbf{x}_1,\mathbf{x}_2)-\log p(\mathbf{x}_1,\mathbf{x}_2|\mathbf{\tilde{y}}_1,\mathbf{\tilde{y}}_2,\mathbf{\tilde{z}}_1,\mathbf{\tilde{z}}_2)-\log p(\mathbf{\tilde{y}}_1,\mathbf{\tilde{y}}_2,\mathbf{\tilde{z}}_1,\mathbf{\tilde{z}}_2)\right\}+\overset{\text{\small const}}{\boxed{\mathbb{E}_{p_{\mathbf{x}_1,\mathbf{x}_2}}
        \log p_{\mathbf{x}_1,\mathbf{x}_2}(\mathbf{x}_1,\mathbf{x}_2)}} \nonumber\\
        &\overset{(b)}{=}\!\mathbb{E}_{p_{\mathbf{x}_1,\mathbf{x}_2}}\mathbb{E}_{q_{\mathbf{\tilde{y}}_1,\mathbf{\tilde{y}}_2,\mathbf{\tilde{z}}_1,\mathbf{\tilde{z}}_2|\mathbf{x}_1,\mathbf{x}_2}}\! \left\{ \log q(\mathbf{\tilde{y}}_1,\mathbf{\tilde{y}}_2,\mathbf{\tilde{z}}_1,\mathbf{\tilde{z}}_2|\mathbf{x}_1,\mathbf{x}_2)-\log p(\mathbf{x}_1,\mathbf{x}_2|\mathbf{\tilde{y}}_1,\mathbf{\tilde{y}}_2,\mathbf{\tilde{z}}_1,\mathbf{\tilde{z}}_2)-\log p(\mathbf{\tilde{y}}_1,\mathbf{\tilde{y}}_2|\mathbf{\tilde{z}}_1,\mathbf{\tilde{z}}_2)-\log p(\mathbf{\tilde{z}}_1,\mathbf{\tilde{z}}_2) \right\}+\text{\small const}\label{sub}\\
        &\overset{(c)}{=}\!\overset{=0}{\boxed{\mathbb{E}_{p_{\mathbf{x}_1,\mathbf{x}_2}}\mathbb{E}_{q_{\mathbf{\tilde{y}}_1,\mathbf{\tilde{y}}_2,\mathbf{\tilde{z}}_1,\mathbf{\tilde{z}}_2|\mathbf{x}_1,\mathbf{x}_2}}\!  \log q(\mathbf{\tilde{y}}_1,\mathbf{\tilde{y}}_2,\mathbf{\tilde{z}}_1,\mathbf{\tilde{z}}_2|\mathbf{x}_1,\mathbf{x}_2)}}\nonumber
        \\&\ \ \ \ \ +\mathbb{E}_{p_{\mathbf{x}_1,\mathbf{x}_2}}\mathbb{E}_{q_{\mathbf{\tilde{y}}_1,\mathbf{\tilde{y}}_2,\mathbf{\tilde{z}}_1,\mathbf{\tilde{z}}_2|\mathbf{x}_1,\mathbf{x}_2}}\left\{-\log p(\mathbf{x}_1,\mathbf{x}_2|\mathbf{\tilde{y}}_1,\mathbf{\tilde{y}}_2,\mathbf{\tilde{z}}_1,\mathbf{\tilde{z}}_2)-\log p(\mathbf{\tilde{y}}_1|\mathbf{\tilde{z}}_1)-\log p(\mathbf{\tilde{y}}_2|\mathbf{\tilde{z}}_2)-\log p(\mathbf{\tilde{z}}_1,\mathbf{\tilde{z}}_2) \right\}+\text{\small const}\nonumber\\
        &=\!\mathbb{E}_{p_{\mathbf{x}_1,\mathbf{x}_2}}\mathbb{E}_{q_{\mathbf{\tilde{y}}_1,\mathbf{\tilde{y}}_2,\mathbf{\tilde{z}}_1,\mathbf{\tilde{z}}_2|\mathbf{x}_1,\mathbf{x}_2}}\! \left\{ -\log p(\mathbf{x}_1,\mathbf{x}_2|\mathbf{\tilde{y}}_1,\mathbf{\tilde{y}}_2,\mathbf{\tilde{z}}_1,\mathbf{\tilde{z}}_2)-\log p(\mathbf{\tilde{y}}_1|\mathbf{\tilde{z}}_1)-\log p(\mathbf{\tilde{y}}_2|\mathbf{\tilde{z}}_2)-\log p(\mathbf{\tilde{z}}_1,\mathbf{\tilde{z}}_2) \right\}+\text{\small const}
        \label{expansion ntsc}
    \end{align}
    \begin{equation}
    q(\mathbf{\tilde{y}}_1,\mathbf{\tilde{y}}_2,\mathbf{\tilde{z}}_1,\mathbf{\tilde{z}}_2|\mathbf{x}_1,\mathbf{x}_2)=\prod_k\mathcal{U}(\tilde{y}_1^k|y_1^k-\frac{1}{2},y_1^k+\frac{1}{2})\prod_j\mathcal{U}(\tilde{y}_2^j|y_2^j-\frac{1}{2},y_2^j+\frac{1}{2})\prod_m\mathcal{U}(z_1^m|z_1^m-\frac{1}{2},z_1^m+\frac{1}{2})\prod_n\mathcal{U}(\tilde{z}_2^n|z_2^n-\frac{1}{2},z_2^n+\frac{1}{2})
    \label{q ntsc}
    \end{equation}
    \hrulefill
    \end{small}
    \vspace{-0.3cm}
\end{figure*}

\begin{figure}[t]
    \centering
    \includegraphics[width=0.48\textwidth]{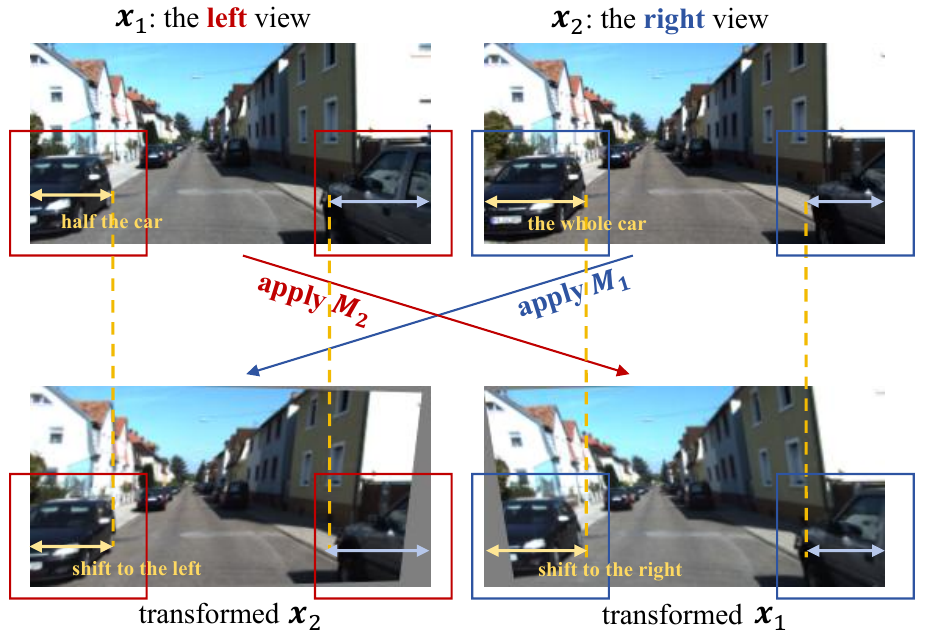}
    \caption{Visualization of the effect of the transformation module.}
    \label{visualization}
    \vspace{-0.3cm}
\end{figure}

In this subsection, we conduct an ablation study to assess the effectiveness of the transformation module.
Fig.~\ref{ablation t} presents the PSNR performance of the D-NTSCC scheme with and without the transformation module at SNR $=$ 2 dB.
Note that ``w/o $\mathcal{T}(\cdot)$'' refers to the case where an identical transformation is applied.
Across different transmission rates, the transformation module consistently improves performance, especially at higher rates.
For example, at $r=0.05$, a PSNR gain of 0.15 dB can be observed.
These results show that by aligning the feature maps, the transformation module indeed facilitates the exploitation of source correlation.

We further apply the learned perspective transformations to the raw images to visualize their effect on the corresponding feature maps. The results are shown in Fig.~\ref{visualization}.
It can be observed that $\mathbf{M}_2$ translates and rotates the left view $\mathbf{x}_1$ to resemble the right view, while $\mathbf{M}_1$ similarly transforms the right view $\mathbf{x}_2$ to $\mathbf{x}_1$.
The visualization confirms the effectiveness of the transformation module in aligning each feature map with the other, which enhances the utilization of side information.

\subsection{Computational Complexity}


\begin{table}[t]
\renewcommand{\arraystretch}{1.4}
\centering
\caption{Computational complexity of different methods.}
\begin{tabular}{l|ccc}
\hline
                                           & Method          & FLOPs (G) & Params (M) \\ \hline
\multicolumn{1}{c|}{\multirow{3}{*}{SSCC}} & D-NTSC          & 67.16  & 54.89  \\
\multicolumn{1}{c|}{}                      & Global Training & 26.58  & 25.40   \\
\multicolumn{1}{c|}{}                      & LDMIC           & 19.53  & 19.91  \\ \hline
\multirow{4}{*}{JSCC}                      & D-NTSCC         & 69.07  & 56.66  \\
                                           & NTSCC           & 28.35  & 32.31  \\
                                           & D-DJSCC         & 60.79  & 13.35  \\
                                           & DHF-JSCC        & 163.68 & 18.36  \\ \hline
\end{tabular}
\label{computation}
\vspace{-0.5cm}
\end{table}

Table \ref{computation} compares the computational complexity of different methods. 
D-NTSC outperforms Global Training and LDMIC in terms of performance, but it comes at the cost of increased computational complexity.
Similarly, D-NTSCC also requires more memory.
In terms of floating point operations (FLOPs), D-NTSCC increases computational complexity by 13.6$\%$ compared to D-DJSCC, but only requires 42.2$\%$ of the computational complexity of DHF-JSCC.

\section{Conclusion}
This paper proposes a general NTC-based approach for distributed image semantic communication, applicable to both SSCC and JSCC paradigms. 
This approach explicitly models source correlation through joint probabilistic modeling and spatial transformation, enabling adaptive rate allocation and improved correlation utilization. 
We implement this framework as D-NTSC for SSCC and D-NTSCC for JSCC, both built on Swin Transformers to enhance feature extraction and correlation exploitation. 
Variational inference is used to derive principled loss functions for joint optimization of encoding, decoding, and entropy modeling.
Experiments demonstrate that both schemes outperform existing distributed baselines, highlighting their potential for distributed semantic communications.

\begin{figure*}
    \begin{small}
    \begin{align}
        &\ \ \ \ \ \mathbb{E}_{p_{\mathbf{x}_1,\mathbf{x}_2}}D_{KL}[q(\mathbf{\hat{s}}_1,\mathbf{\hat{s}}_2,\mathbf{\tilde{z}}_1,\mathbf{\tilde{z}}_2|\mathbf{x}_1,\mathbf{x}_2)||p(\mathbf{\hat{s}}_1,\mathbf{\hat{s}}_2,\mathbf{\tilde{z}}_1,\mathbf{\tilde{z}}_2|\mathbf{x}_1,\mathbf{x}_2)]\nonumber\\
        &\overset{(a)}{=}\!\mathbb{E}_{p_{\mathbf{x}_1,\mathbf{x}_2}}\mathbb{E}_{q_{\mathbf{\hat{s}}_1,\mathbf{\hat{s}}_2,\mathbf{\tilde{z}}_1,\mathbf{\tilde{z}}_2|\mathbf{x}_1,\mathbf{x}_2}}\! \left\{ \log q(\mathbf{\hat{s}}_1,\mathbf{\hat{s}}_2,\mathbf{\tilde{z}}_1,\mathbf{\tilde{z}}_2|\mathbf{x}_1,\mathbf{x}_2)-\log p(\mathbf{x}_1,\mathbf{x}_2|\mathbf{\hat{s}}_1,\mathbf{\hat{s}}_2,\mathbf{\tilde{z}}_1,\mathbf{\tilde{z}}_2)-\log p(\mathbf{\hat{s}}_1,\mathbf{\hat{s}}_2|\mathbf{\tilde{z}}_1,\mathbf{\tilde{z}}_2)-\log p(\mathbf{\tilde{z}}_1,\mathbf{\tilde{z}}_2) \right\}+\text{\small const}\nonumber\\
        &\overset{(b)}{=}\!\overset{\text{\small const}'}{\boxed{\mathbb{E}_{p_{\mathbf{x}_1,\mathbf{x}_2}}\mathbb{E}_{q_{\mathbf{\hat{s}}_1,\mathbf{\hat{s}}_2,\mathbf{\tilde{z}}_1,\mathbf{\tilde{z}}_2|\mathbf{x}_1,\mathbf{x}_2}}\!  \log q(\mathbf{\hat{s}}_1,\mathbf{\hat{s}}_2,\mathbf{\tilde{z}}_1,\mathbf{\tilde{z}}_2|\mathbf{x}_1,\mathbf{x}_2)}}\nonumber
        \\&\ \ \ \ \ +\mathbb{E}_{p_{\mathbf{x}_1,\mathbf{x}_2}}\mathbb{E}_{q_{\mathbf{\hat{s}}_1,\mathbf{\hat{s}}_2,\mathbf{\tilde{z}}_1,\mathbf{\tilde{z}}_2|\mathbf{x}_1,\mathbf{x}_2}}\left\{-\log p(\mathbf{x}_1,\mathbf{x}_2|\mathbf{\hat{s}}_1,\mathbf{\hat{s}}_2,\mathbf{\tilde{z}}_1,\mathbf{\tilde{z}}_2)-\log p(\mathbf{\hat{s}}_1|\mathbf{\tilde{z}}_1)-\log p(\mathbf{\hat{s}}_2|\mathbf{\tilde{z}}_2)-\log p(\mathbf{\tilde{z}}_1,\mathbf{\tilde{z}}_2) \right\}+\text{\small const}\nonumber\\
        &\overset{(c)}{=}\!\mathbb{E}_{p_{\mathbf{x}_1,\mathbf{x}_2}}\mathbb{E}_{q_{\mathbf{\hat{s}}_1,\mathbf{\hat{s}}_2,\mathbf{\tilde{z}}_1,\mathbf{\tilde{z}}_2|\mathbf{x}_1,\mathbf{x}_2}}\! \left\{ -\log p(\mathbf{x}_1,\mathbf{x}_2|\mathbf{\hat{s}}_1,\mathbf{\hat{s}}_2,\mathbf{\tilde{z}}_1,\mathbf{\tilde{z}}_2)-\log p(\mathbf{\hat{s}}_1|\mathbf{\tilde{z}}_1)-\log p(\mathbf{\hat{s}}_2|\mathbf{\tilde{z}}_2)-\log p(\mathbf{\tilde{z}}_1,\mathbf{\tilde{z}}_2) \right\}+\text{\small const}''
        \label{expansion}
    \end{align}
    \begin{equation}
    q(\mathbf{\hat{s}}_1,\mathbf{\hat{s}}_2,\mathbf{\tilde{z}}_1,\mathbf{\tilde{z}}_2|\mathbf{x}_1,\mathbf{x}_2)=\prod_k\mathcal{N}(\hat{s}_1^k|s_1^k,\epsilon_1^2)\prod_j\mathcal{N}(\hat{s}_2^j|s_2^j,\epsilon_2^2)\prod_m\mathcal{U}(z_1^m|z_1^m-\frac{1}{2},z_1^m+\frac{1}{2})\prod_n\mathcal{U}(\tilde{z}_2^n|z_2^n-\frac{1}{2},z_2^n+\frac{1}{2})
    \label{q}
    \end{equation}
    \hrulefill
    \end{small}
    \vspace{-0.3cm}
\end{figure*}

\begin{appendices}

\section{Proof of Proposition \ref{prop 1}}

To prove proposition 1, we begin by expanding the KL divergence as shown in \eqref{expansion ntsc}.
Equation (a) follows directly from the definition of the KL divergence.
The last term in (a) is a constant, as the source probability remains fixed.
Equation (b) applies the multiplication rule of probability.
Equation (c) holds due to the Markov chain $\mathbf{\tilde{y}}_1-\mathbf{\tilde{z}}_1-\mathbf{\tilde{z}}_2-\mathbf{\tilde{y}}_2$, which leads to the factorization $p(\mathbf{\tilde{y}}_1,\mathbf{\tilde{y}}_2|\mathbf{\tilde{z}}_1,\mathbf{\tilde{z}}_2)=p(\mathbf{\tilde{y}}_1|\mathbf{\tilde{z}}_1)p(\mathbf{\tilde{y}}_2|\mathbf{\tilde{z}}_2)$.
Furthermore, the first term in (c) evaluates to zero, since given $\mathbf{x}_1$ and $\mathbf{x}_2$, $q(\mathbf{\tilde{y}}_1,\mathbf{\tilde{y}}_2,\mathbf{\tilde{z}}_1,\mathbf{\tilde{z}}_2|\mathbf{x}_1,\mathbf{x}_2)$ reduces to a product of uniform distributions, as shown in \eqref{q ntsc}, which completes the proof.

\section{Proof of Proposition \ref{prop 2}}

The proof of Proposition 2 follows a similar idea to that of Proposition 1.
Simply by substituting $\mathbf{\tilde{y}}_1$ with $\mathbf{\hat{s}}_1$ and $\mathbf{\tilde{y}}_2$ with $\mathbf{\hat{s}}_2$ in \eqref{sub}, we can obtain Equation (a).
Equation (b) follows from the Markov chain $\mathbf{\hat{s}}_1-\mathbf{\tilde{z}}_1-\mathbf{\tilde{z}}_2-\mathbf{\hat{s}}_2$, which implies the factorization $p(\mathbf{\hat{s}}_1,\mathbf{\hat{s}}_2|\mathbf{\tilde{z}}_1,\mathbf{\tilde{z}}_2)=p(\mathbf{\hat{s}}_1|\mathbf{\tilde{z}}_1)p(\mathbf{\hat{s}}_2|\mathbf{\tilde{z}}_2)$.
Moreover, the first term in (b) evaluates to a constant, since given $\mathbf{x}_1$ and $\mathbf{x}_2$, $q(\mathbf{\hat{s}}_1,\mathbf{\hat{s}}_2,\mathbf{\tilde{z}}_1,\mathbf{\tilde{z}}_2|\mathbf{x}_1,\mathbf{x}_2)$ reduces to the product of known distributions, as shown in \eqref{q}.
This leads to the final expression in Equation (c), where $\text{const}''=\text{const}+\text{const}'$, thus completing the proof.

\end{appendices}





\bibliographystyle{IEEEtran}
\bibliography{dntscc_j_ver}{}

\end{document}